\documentclass[useAMS,usenatbib]{mn2e}
\usepackage{amsmath}
\usepackage{amssymb}
\usepackage{multirow}
\usepackage{graphicx}
\usepackage{epstopdf}

\usepackage{color}

\def\simlt{\lower.5ex\hbox{$\; \buildrel < \over \sim \;$}}
\def\simgt{\lower.5ex\hbox{$\; \buildrel > \over \sim \;$}}

\newcommand{\bd}{\begin{displaymath}}
\newcommand{\ed}{\end{displaymath}}
\newcommand{\be}{\begin{equation}}
\newcommand{\ee}{\end{equation}}
\newcommand{\beqa}{\begin{eqnarray}}
\newcommand{\eeqa}{\end{eqnarray}}

\title[21-cm Signature with Extra Radio Background] 
{Signature of Excess Radio Background in the 21-cm Global Signal and Power Spectrum} 
\author[Fialkov \& Barkana] {Anastasia Fialkov$^{1,2,3,4, 5}$\thanks{E-mail:
    anastasia.fialkov@gmail.com},  
  Rennan Barkana$^{6}$ \\
  $^{1}$ Harvard-Smithsonian Center for Astrophysics, 60 Garden Street, Cambridge, MA 02138, USA\\
$^{2}$ Department of Physics, The University of Tokyo, 7-3-1 Hongo, Bunkyo, Tokyo 113-0033, Japan \\
  $^{3}$ Institute of Astronomy, University of Cambridge, Madingley Road, Cambridge CB3 0HA, UK\\
$^{4}$ Kavli Institute for Cosmology, University of Cambridge, Madingley Road, Cambridge CB3 0HA, UK \\
$^{5}$ Department of Physics and Astronomy, University of Sussex, Falmer,  Brighton BN1 9QH, UK \\
  $^{6}$ Raymond and Beverly Sackler School of Physics and Astronomy,    Tel Aviv University, Tel Aviv 69978, Israel}

\begin{document}
\pagerange{\pageref{firstpage}--\pageref{lastpage}} \pubyear{2015}
\maketitle

\label{firstpage}

\begin{abstract}

The recent tentative detection by the EDGES Low-Band of the hydrogen 21-cm line from cosmic dawn, if confirmed, is the first ever signature observed from the epoch of primordial star formation. However, the magnitude and the shape of this signal are incompatible with standard astrophysical predictions, requiring either colder than expected gas, or an excess radio background above the Cosmic Microwave Background (CMB) radiation. In this paper we explore the latter option, investigating possible 21-cm signals in models with standard astrophysics to which a phenomenological excess radio background was added. Assuming uniform radiation with a synchrotron-like spectrum  and redshift-independent amplitude existing throughout cosmic history, we show that such a radio background, in addition to explaining the EDGES detection, enhances the power spectrum. The signal during cosmic dawn and reionization strongly depends on both the intensity of the background and the astrophysical parameters. We verify the broad agreement of our models with the detected feature, including additional constraints from the EDGES High-Band, high-redshift quasars, and observational limits on the extragalactic radio background. The data imply a lower limit on the star formation efficiency of 2.8\%, an upper limit on the minimum mass of star-forming halos of M$_{\rm h} =$ 10$^9$ M$_\odot$ at z = 17, and a lower limit on the excess background of 1.9 times the CMB at 78 MHz. The properties of X-ray sources are unconstrained by the data. We also show that during the dark ages, such a radio background saturates the 21-cm signal, imprinting a unique signature.

\end{abstract}


\section{Introduction}
\label{Sec:Intro}

The evolution history of the neutral Universe in the first few hundred million years after the Big Bang and  the process of reionization remain poorly constrained. Theoretical modeling bridges between the observations of the two dimensional surface of recombination at redshift $z\sim 1100$ probed by the Cosmic Microwave Background (CMB) radiation \citep[e.g.,][]{Planck:2018},  and the epoch of  galaxy formation with the most distant detected galaxy located at $z = 11.1$ \citep{Oesch:2016}. Observations of high redshift dusty galaxies indicate an early onset of star formation. Specifically, recent detection of doubly ionized oxygen at redshift  $z = 9.1$  implies a dominant stellar component that formed at $z\sim 15$ \citep[$\sim 250$ million years after the Big Bang,][]{Hashimoto:2018}. The appearance of massive galaxies of M$_{\rm h} \gtrsim 10^9$ M$_\odot$ leads to an efficient reionization. The most recent constraints on the neutral hydrogen fraction, $x_{\rm HI}$, from the damping wings of ULASJ1342+0928, a bright QSO at z = 7.54 \citep{Banados:2018}, show that the Universe was only $21^{+17}_{-19}$\%  neutral  at that epoch \citep[at 68\% confidence,][]{Greig:2018}; while the upper limit on the neutral fraction at redshift $5.9$ is $x_{\rm HI}\lesssim 6+5$\%  \citep[at 68\% confidence,][]{Greig:2017}. 

Recently, the first detection of the sky-averaged (global) 21-cm signal of neutral hydrogen from $z\sim 13-27$ has been claimed based on two years of observations with the Low-Band antenna of the Experiment to Detect the Global EoR Signature (EDGES) in the $50-100$ MHz frequency range \citep{Bowman:2018}. If confirmed, this is the only existing observation  from the intermediate redshift range and is the first observational evidence of the primordial star formation at $z\sim 20$ ($\sim 180$ million years after the Big Bang) and early X-ray heating. This detection is consistent with non-detection of the 21-cm signal at higher frequencies by the EDGES High-Band \citep{Monsalve:2017, Monsalve:2018, Monsalve:2018b} and the Shaped  Antenna measurement of the background RAdio Spectrum2  \citep[SARAS2,][]{Singh:2017, Singh:2018} which indicates that the signal does not vary strongly at redshifts  $z\sim 6-15$. In the framework of standard astrophysical modeling, the data requires star formation in small dark matter halos \citep{Monsalve:2018b} and rules out extremely non-efficient X-ray heating sources \citep{Monsalve:2018b, Singh:2017, Singh:2018}.

The reported cosmological 21-cm signal is centered at $z = 17.2$ (which corresponds to $\nu  = 78.2$ MHz), and features an absorption trough of $T_{21} = -500^{+200}_{-500}$ mK, where the error corresponds to 99\% confidence including both thermal and systematic noise. The depth of the feature is at least twice as strong as predicted in standard astrophysical scenarios (based on the assumption of $\Lambda$CDM cosmology and hierarchical structure formation) where the strongest possible feature at $z = 17$ is $-209$ mK,  assuming the CMB as the background radiation with $T_{\rm CMB} = 49.5$ K and the coldest possible temperature of the intergalactic medium (IGM) of $\sim 7.2$ K obtained in the absence of X-ray heating sources. The observed $T_{21} < - 300$ mK requires either  the gas to be much colder (around 5.2  K) or the background radiation to be much stronger (around  67.2  K), which is hard to explain by astrophysics alone. Finally, no viable model has been proposed to explain the flatness of the best-fit profile (though the precise shape of the profile may be sensitive to the systematic errors and to the assumed form of the fitted model). Observational effort is underway to test the EDGES Low-Band signature with global signal instruments including Large Aperture Experiment to Detect the Dark Ages \citep[LEDA,][]{Price:2018}, SARAS3 \citep[variation of SARAS2,][optimized in the 50-100 MHz frequency range]{Singh:2017}, Probing Radio Intensity at high-Z from Marion \citep[PRIZM,][]{Philip:2018}; while operating low frequency interferometers (e.g., LOFAR) might be sensitive to the 21-cm power spectrum if it is enhanced accordingly \citep[e.g., ][]{Fialkov:2018}.   

The frequency of the detected feature can be related to the formation of the first stars which couple the spin temperature to the temperature of the gas via   absorption and re-emission of Ly-$\alpha$ photons \citep[the Wouthuysen-Field (WF) effect, ][]{Wouthuysen:1952, Field:1958}. The narrowness of the feature indicates a relatively rapid onset of X-ray heating. Assuming an excess radio background created by high-redshift star forming galaxies and by trying to match the observed profile,  \citet{Mirocha:2018} find that the location of the absorption trough implies efficient star formation in $10^8-10^{10}$ M$_\odot$ halos and efficient X-ray sources with $f_X>10$; \citet{Kaurov:2018} find that (in the absence of an excess radio background) rapid Ly-$\alpha$ coupling is required to reproduce the steepness of the best-fit profile indicating M$_{\rm h}\gtrsim 10^9$ M$_\odot$; however, \citet{Schauer:2018} show that smaller mass halos are needed to ensure efficient Ly-$\alpha$  coupling at $z\sim17$. 

To explain the  depth of the reported feature exotic mechanisms have to be added to the standard picture. The amplitude of the observed 21-cm feature can be explained by non-gravitational interaction between dark matter and  baryons, e.g., via Rutherford-like scattering which could drain access energy from the gas lowering its kinetic temperature \citep{Tashiro:2014, Dvorkin:2014, Munoz:2015,  Barkana:2018, Slatyer:2018}. Even though this scenario is strongly constrained by observations \citep{Barkana:2018b, Berlin:2018, Kovetz:2018, Munoz:2018}, it is not completely ruled out and is still plausible for a narrow range of parameters including the dark matter mass, electric charge of dark matter particle and cross-section \citep[e.g.,][]{Kovetz:2018}. A smoking gun signature of the baryon-dark matter (b-dm) scattering with a velocity-dependent cross-section, first proposed by \citet{Barkana:2018} as a solution to the EDGES Low-Band anomaly, was shown to be an enhanced pattern of Baryon Acoustic Oscillations (BAO) in the 21-cm power spectrum \citep{Fialkov:2018, Munoz:2018b}. The power spectrum itself is boosted by as much as three orders of magnitude which renders  the fluctuations detectable by telescopes such as the low band antennas of LOFAR  (private communication with L. Koopmans).

An excess radio background at the rest-frame 1.42 GHz at $z=17$ would also  explain the large contrast observed between the background temperature and the spin temperature of the 21-cm transition \citep{Bowman:2018, Feng:2018}. Astrophysical sources such as accreting supermassive black holes \citep{Biermann:2014, EwallWice:2018} or supernovae \citep{Jana:2018, Mirocha:2018}  could generate the extra radio background via synchrotron emission produced by electrons accelerated in a magnetic field. However, both types of sources would need to be a thousand times more efficient in producing synchrotron radiation than their low-redshift counterparts. Moreover, inverse-Compton cooling off the CMB photons was shown to greatly reduce the efficiency of the radio sources at high redshifts \citep{Oh:2001, Ghisellini:2014, Sharma:2018}. An extra radio background can also be created by more exotic processes in which case its intensity is not related to the star formation history. Examples include radiative decay of relic neutrinos into sterile neutrinos \citep{Chianese:2018}, models in which additional radiation at 1.42 GHz is injected into the gas via light dark matter decays \citep{Fraser:2018, Pospelov:2018} and superconducting cosmic strings \citep{Brandenberger:2019}.  Interestingly, evidence of excess radio background above the CMB at low radio frequencies was detected by ARCADE2 at $3-90$ GHz \citep{Fixsen:2011} and recently confirmed by LWA1  at $40-80$ MHz \citep{Dowell:2018}. LWA1 measurements are consistent with excess  background which can be fitted  by a power law with a spectral index of$ ~- 2.58\pm 0.05$ and a temperature of $603^{+102}_{ - 92}$ mK at 1.42 GHz. However, it is still not clear what part of the observed excess is extragalactic  \citep{Subrahmanyan:2013}.

Finally, modification of the thermal history can also be a result of earlier thermal decoupling of baryons from the CMB which can occur as a result of early dark energy \citep[however, such scenarios are ruled out by other observations][]{Hill:2018} or due to an imbalance between the proton and electron number densities \citep{Falkowski:2018}. Interacting dark energy could modify the evolution of the Hubble parameter and, as a result, affect the 21-cm brightness temperature \citep{Costa:2018}; however, in this case the change in the Hubble parameter at $z\sim 20$ would be too large.

In this paper we use hybrid computational methods to explore the 21-cm signal in models with standard astrophysics \citep[e.g.,][]{Cohen:2017} to which an excess uniform radio background  of redshift-independent amplitude and synchrotron-like spectrum was added.  Although this observationally inspired spectrum does not fully represent the above-mentioned  realistic scenarios which either generate background with redshift-dependent amplitude and/or a more complicated spectral dependence, it provides a useful guideline.  Such a contribution is usually not considered in 21-cm studies in the literature, where the standard CMB is used as the background  \citep[e.g.,][]{Mesinger:2011}. The paper is organized as follows: we outline the methodology and show a few examples of the global signals and the power spectra in Section \ref{Sec:Methods} highlighting the implications of the excess radio background.  To explore the dependence of the results on the shape of the radio spectrum,  we show several cases with different  spectral indexes. We also show that, although the background itself is uniform, it affects the overall level of fluctuations  as well as the global signal.   In Section \ref{Sec:params} we use the reported best-fit EDGES Low-Band profile, together with the limits on the 21-cm signal from the EDGES High-Band and measurements of $x_{\rm HI}$, to put constraints on the joint parameter space of standard astrophysics as well as the amplitude of the extra radio background. We discuss the smoking gun signatures of this model and compare them to the signatures of the models with b-dm scattering in Section \ref{Sec:disc}. Finally, we conclude in Section \ref{Sec:sum}.

\section{Modeling}
\label{Sec:Methods}

\subsection{Thermal and Ionization Histories}

Theory predicts that, as the Universe evolves, astrophysical processes drive major changes in the global thermal and ionization histories. Right after recombination residual electrons were efficient  in coupling the gas temperature to that of the background radiation. As a result, gas and radiation were cooling together at the rate $\propto (1+z)$. However, as a result of the expansion of the Universe, the coupling became inefficient and, at around $z\sim200$, the cooling rate of the gas increased to $\propto (1+z)^2$. At that epoch gas was cooling faster than the background. This period in cosmic history, in the absence of luminous astrophysical sources, is called the dark ages and is characterized by a cold and neutral intergalactic medium.  The emergence of the first stellar and first X-ray populations changed  the environment at high redshifts and marked the onset of cosmic dawn. During this epoch stars produce copious Ly-$\alpha$ radiation, and the gas is reheated  by the injected X-ray photons which also mildly re-ionize it. The heating rate and the character of fluctuations in the gas temperature  depend on the properties of X-ray sources such as the spectral energy distribution (SED) and bolometric luminosity \citep[e.g.,][]{Fialkov:2014, Fialkov:2014b, Pacucci:2014}. The process of reionization happens slower than heating as the first small galaxies are thought to be inefficient in ionizing the gas.  The appearance of massive galaxies of M$_{\rm h} \gtrsim 10^9$ M$_\odot$ leads to a more efficient reionization which, as observations suggest, starts not long before redshift 7.5 and terminates at $z\sim 6$ \citep[e.g.,][]{Greig:2017, Greig:2018}.  The 21-cm signal  is thought to be one of the most promising probes of the global thermal and ionization histories. 

\subsection{21-cm Signal}
\label{Sec:21cm}

The 21-cm signal is produced by neutral hydrogen atoms in the IGM. The Hyper-fine splitting of the lowest hydrogen energy level gives rise to the rest-frame $\nu_{21} = 1.42$ GHz radio signal with the equivalent wavelength of 21 cm \citep[see][for a recent review]{Barkana:2016}.  The cumulative signal of neutral IGM observed against the background radiation depends on the processes of cosmic heating and ionization and, to the leading order, scales as $T_{21} \propto x_{\rm HI}\left(1-T_{\rm rad}T_{\rm S}^{-1}\right)$ where $T_{\rm rad}$  is the brightness temperature of the background radiation  at 1.42 GHz and $T_{\rm S}$ is the spin temperature of the transition which at cosmic dawn is close to the kinetic temperature of the gas, $T_{\rm K}$. Owing to its dependence on the underlying astrophysics and cosmology, this signal is a powerful tool to characterize the formation and the evolution of the first populations of astrophysical sources and, potentially, properties of dark matter, across cosmic time. 

 The brightness temperature of the 21-cm transition of neutral hydrogen of spin temperature $T_{\rm S}$ observed against  background radiation of brightness temperature $T_{\rm rad}$ is given by:
\begin{equation}
T_{21}  = \frac{T_{\rm S}-T_{\rm rad}}{1+z}\left(1-e^{-\tau_{21}}\right).
\label{Eq:T}  
\end{equation}
The 21-cm optical depth is 
\begin{equation}
\tau_{21} = \frac{3h_{\rm pl}A_{10}c \lambda_{21}^2n_{\rm H}}{32\pi k_{B} T_{\rm S} (1+z) dv/dr},
\end{equation}
where  $dv/dr = H(z)/(1+z)$ is the gradient of the line of sight component of the comoving velocity field, $H(z)$ is  the Hubble constant, and $n_{\rm H}$ is the neutral hydrogen number density which depends on the ionization history driven by both ultraviolet and X-ray photons. Usually, a linearized version of Eq. \ref{Eq:T} is used. However, in cases of cold gas temperature the linear expression is not necessarily a good approximation.  The spin temperature encodes complex astrophysical dependence
\begin{equation}
T_{\rm S} = \frac{x_{\rm rad}+x_{\rm tot}}{x_{\rm rad}T_{\rm rad}^{-1}+x_{\rm tot}T_{\rm K}^{-1}},
\end{equation}
where 
\begin{equation}
x_{\rm rad}=(1-e^{-\tau_{21}})/\tau_{21}
\label{Eq:xrad}
\end{equation}
 depends on $T_{\rm S}$ \citep{Venumadhav:2018}. To solve for $x_{\rm rad}$ and $T_{\rm S}$ we iterate, starting from the value $x_{\rm rad} =1$, until the values of $x_{\rm rad}$ and $T_{\rm S}$ converge.  We follow \citet{Barkana:2016} to calculate $x_{\rm tot} = x_C +x_{\alpha,{\rm eff}}$ \citep[Eqs. 46 and 48][]{Barkana:2016}, where $x_{\alpha,{\rm eff}} \sim x_\alpha$  \citep[Eq. 57 of][]{Barkana:2016} is the WF coupling coefficient 
 \begin{equation}
 x_\alpha = \frac{4P_\alpha}{27 A_{10}}\frac{T_*}{T_{\rm rad}}
 \label{eq:xa}
 \end{equation}
 with $P_\alpha$ being the total rate (per atom) at which Ly-$\alpha$ photons are scattered within the gas and $T_*$ is the effective temperature of the 21-cm transition (0.068 K). $x_C$ is the collisional coupling coefficient 
\begin{equation}
x_C = \frac{n_i \kappa^i_{10}}{A_{10}}\frac{T_*}{T_{\rm rad}}
\label{eq:xc}
\end{equation} 
 where $\kappa^i_{10}$  is the rate coefficient for spin de-excitation in collisions with the
species of type $i$ of density $n_i$ where we sum over species $i$.  

Usually, the background radiation is assumed to be the CMB, $T_{\rm rad} = T_{\rm CMB}(1+z)$, where  $T_{\rm CMB}$ is the CMB temperature today, 2.725 K. Here we add a phenomenological uniform excess inspired by the ARCADE2 and LWA1 observations. The total radio background has the form
\begin{equation}
T_{\rm rad} = T_{\rm CMB}(1+z)\left[1+A_{\rm r}\left(\frac{\nu_{\rm obs}}{78~{\rm MHz}}\right)^{\beta}\right]
\label{Eq:Trad}
\end{equation}
where  $\nu_{\rm obs}$ is the observed frequency,  $A_{\rm r}$ is the amplitude defined relative to the CMB temperature and $\beta$ is the spectral index.

We solve coupled equations for   $T_{\rm K}$ and the free electron fraction, $x_{\rm e}$, taking into account the following heating and cooling terms: adiabatic cooling, Compton heating (which depends on the total radiation energy density dominated by the CMB), the contribution of X-ray sources (described in Section \ref{Sec:astro}),  heating by Ly-$\alpha$ photons as well as radiative heating which was recently pointed out by \citet{Venumadhav:2018}. We account for recombinations and the ionization rate due to X-ray sources.  Ionization by stellar ultraviolet photons is implemented separately using the excursion-set approach \citep{Furlanetto:2004}.

Fully self-consistent  modeling of the 21-cm signal from large cosmological volumes is prohibitively expensive  as it requires one both to simulate the distribution of sources and the intensity of radiative backgrounds that fluctuate on scales of a few hundred comoving Mpc and resolve small scales relevant for star and black hole formation. Here we use hybrid computational methods that tracks the large scale evolution on scales above 3 comoving Mpc while extensively using sub-grid models to account for star formation and the X-ray sources. Our modeling is extensively described in \citet{Visbal:2012, Fialkov:2014, Fialkov:2014b, Cohen:2017} and is partly inspired by the publicly available 21cmFAST \citep{Mesinger:2011}. Compared to our previous work here we improve by: (1) using the non-linear expression for $T_{21}$ (Eq. \ref{Eq:T}), (2) adding the radiative coupling coefficient $x_{\rm rad}$ (Eq. \ref{Eq:xrad}), and (3) accounting for the radiative heating term which  contributes about $\mathcal{O}(10\%)$ of the total heating rate in the scenarios with inefficient X-ray heating \citep{Venumadhav:2018}. We replace the CMB background  radiation by a more generic frequency-dependent radiative background $T_{\rm rad}$, Eq. \ref{Eq:Trad}.  In addition to the straightforward dependence of $T_{21}$ on $T_{\rm rad}$ in Eq. \ref{Eq:T}, the extra radio background affects the calculation of $x_C$ and $x_\alpha$. 

We use this method to calculate the 21-cm signal in a volume of 384$^3$ Mpc$^3$ at redshifts from 300 to 6 covering the dark ages, cosmic dawn and reionization. We extract the global signal and the large-scale power spectrum.

 In this paper we assume a homogeneous radio background. However, if radio sources are distributed inhomogeneously around the sky (e.g., due to the clustering of galaxies), fluctuations in the background, $\delta T_{\rm rad}$, are imprinted. In this case the radio background acts as a  new source of fluctuations in the 21-cm signal of the form 
\begin{equation}
\delta T_{21} = \frac{\partial T_{21}}{\partial T_{\rm rad}}\delta T_{\rm rad} = 
\frac{\tau_{21}}{(1+z)}\frac{-x_{\rm tot}}{\left(1+x_{\rm tot}\right)}\delta T_{\rm rad}
\end{equation} 
where for simplicity we assumed $x_{\rm rad}=1$, small optical depth $\tau_{21}$ and gas temperature $T_{\rm K}$ independent of the radio background. Such low-frequency radio waves would travel far without significant
absorption, so the effective horizon of sources would be set by the
time-retarded decline of star formation, modified by the effect of
redshift as well as the source radio spectrum. Such fluctuations have not been considered before and will be included in our future work.

\subsection{Model Parameters}
\label{Sec:astro}

Astrophysical processes affect the 21-cm signal by driving the spin temperature. We include the following processes \citep[see][for more details]{Cohen:2017}: 
\begin{itemize}
\item The Ly-$\alpha$ photons produced by stellar sources, couple the spin temperature of neutral hydrogen atoms to the kinetic temperature of the gas \citep{Wouthuysen:1952, Field:1958}. Molecular hydrogen cooling allows star formation in small halos of circular velocity $4.2<V_c<16.5$ km s$^{-1}$ (M$_{\rm h}\sim 10^5 -10^7$ M$_\odot$) which are abundant at cosmic dawn. However, the $H_2$ molecule is fragile and star formation in such halos is suppressed by Lyman-Werner (LW) radiation. Additional inhomogeneous suppression is introduced by the relative velocity between dark matter and baryons \citep[$v_{\rm bc}$,][]{Tseliakhovich:2010} which imprints the pattern of BAO in the 21-cm signal \citep{Dalal:2010, Visbal:2012}. Higher mass halos ($V_c>16.5$ km s$^{-1}$) form stars via atomic hydrogen cooling and are sensitive to neither the LW feedback nor to the effect of $v_{\rm bc}$.  The process of star formation is parameterized by the star formation efficiency, $f_*$, which we vary in the range $f_* = 0.028$\% to 50\% and the value of circular velocity which is varied between 4.2 km s$^{-1}$ (molecular hydrogen cooling, corresponding to M$_{\rm h}=6.3\times 10^{5}$ M$_\odot$ at $z=17$) and 100 km s$^{-1}$ (M$_{\rm h}=8.5\times 10^{9}$ M$_\odot$ at $z=17$) where the high values implicitly take into account various chemical and mechanical feedbacks which we do not include explicitly.  
\item X-ray sources re-heat the gas after the period of adiabatic cooling. The X-ray spectrum is modeled as a  power-law with spectral index $\alpha$ which we vary between $-1.5$ and $-1$, lower energy cutoff $\nu_{\rm min}$ in the range $0.1-3$ keV, and normalization constant $f_X$ defined via the ratio of the total X-ray luminosity between $\nu_{\rm min}$ and 100 keV, $L_X$, to the star formation rate, $L_X/SFR = 3\times 10^{40}f_X$ erg s$^{-1}$ M$_\odot^{-1}$ yr. $f_X =1 $ yields $L_X$ normalized to observations of
high-redshift X-ray binaries. We explore the range $f_X = 10^{-6}-10^3$; however, values of $f_X \gtrsim 100$ are unlikely as such a population would saturate the unresolved X-ray background observed by the {\it Chandra}  X-ray Observatory \citep{Fialkov:2017}.    
\item Stellar sources emit ultraviolet radiation which ionizes the gas. Reionization is described by the ionizing efficiency of sources, $\zeta$, which at every $V_c$ is required to give a CMB optical depth $\tau$ in the range between 0.05 and 0.09, in broad agreement with {\it Planck} data \citep[e.g.,][]{Planck:2018}, and the mean free path of ionizing photons, R$_{\rm mfp} = 10-50$ Mpc. The ionizing efficiency has an upper limit of  $\zeta <40000 f_*$ in agreement with stellar models of extreme massive population III stars \citep{Bromm:2001}.
  
\item A uniform radio background is added to the CMB and is parameterized by the amplitude $A_{\rm r}$, calculated at the reference frequency of 78 MHz, and spectral index $\beta$ as discussed in Section \ref{Sec:21cm}.  We vary the value of $A_{\rm r}$ between 0.3 (0.016\% of the CMB at 1.42 GHz) and  2000 (with $A_{\rm r,max} = 418$ at 78 MHz equivalent to the LWA1 limit, a fraction of 0.22 of the CMB at 1.42 GHz) and  assume the fiducial value of $\beta = - 2.6$  in agreement with the observed excess found by LWA1 between $40-80$ MHz.  We also show cases with $\beta = -2$ and $-3$ in Figure 1.
\end{itemize}

\begin{figure*}
\begin{center}
\includegraphics[width=3.5in]{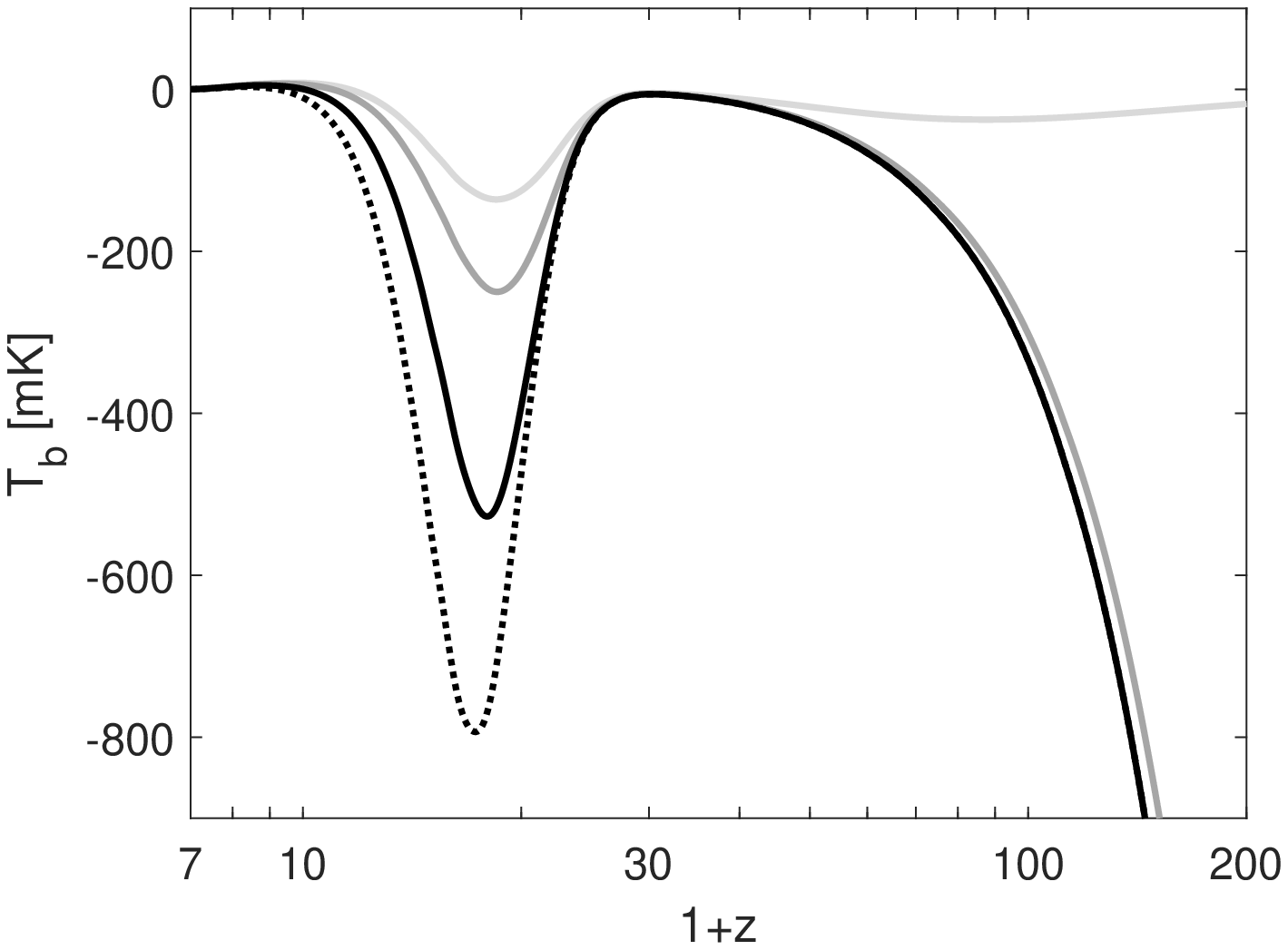}\includegraphics[width=3.5in]{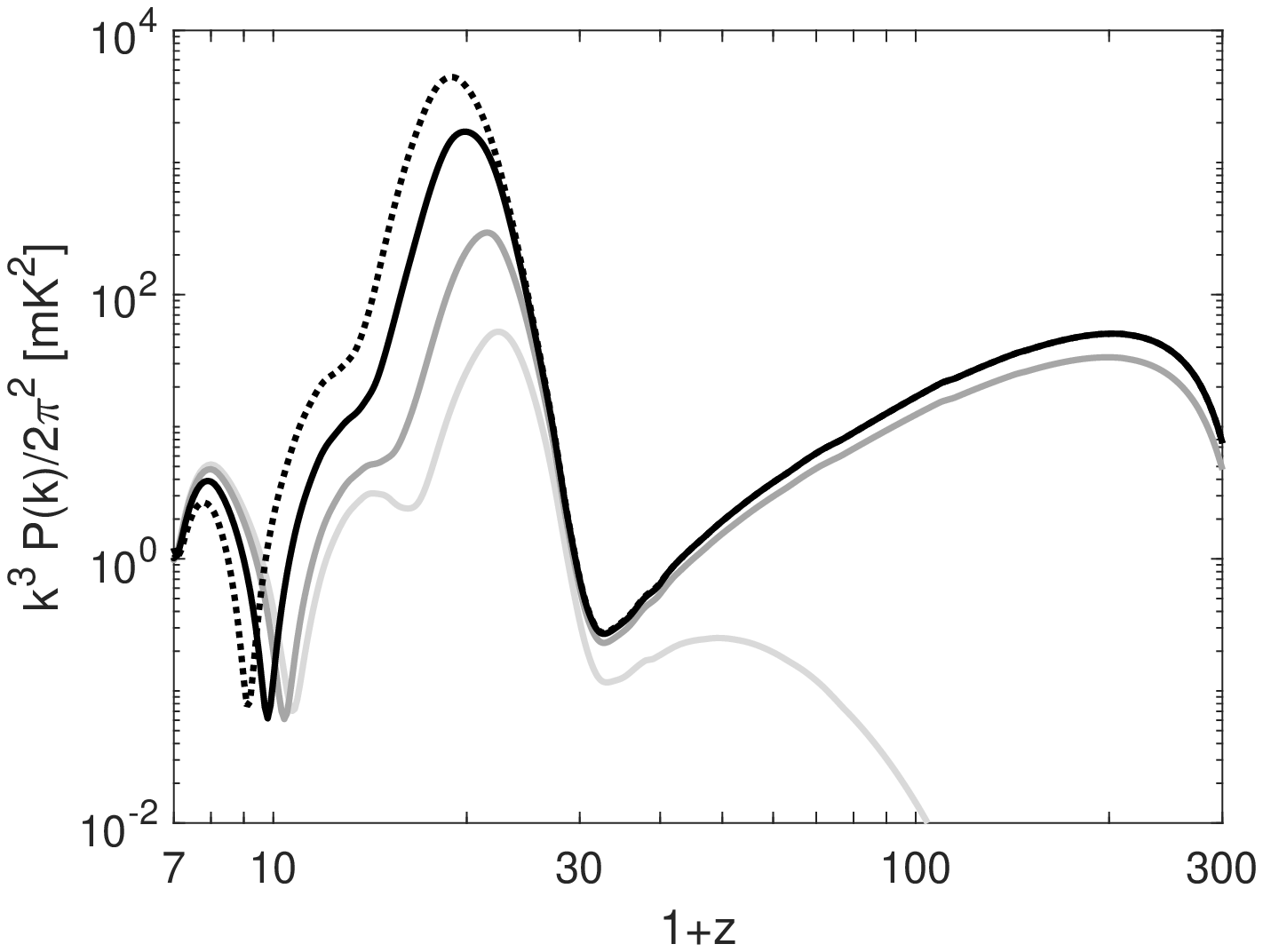}
\caption{Comparison for the global signal (left) and the power spectrum at k  = 0.1 Mpc$^{-1}$ (right). We show a case with  $A_{\rm r} = 0$ (light grey) which is the case of ``standard astrophysics'', and three cases with the extra radio background of  $A_{\rm r} = 1$ and $ \beta=-2$  (dark grey),  $A_{\rm r} = 5.7$ and $ \beta=-2.6$ (solid black), and 
 $A_{\rm r} = 18.4$ and $ \beta=-3$ (dotted black).  All the considered cases assume the same set of astrophysical parameters: $f_* = 3\%$, $f_X = 1$, $V_c = 16.5$ km s$^{-1}$, R$_{\rm mfp} = 30$ Mpc, $\zeta = 14$, $\alpha = -1.3$, $\nu_{\rm min} =1$ keV (hard X-ray SED). } 
\end{center}
\label{fig:signals}
\end{figure*} 

\subsection{Case study}
\label{Sec:case}

We demonstrate the effect of the extra radio background by comparing  the 21-cm global signal and power spectrum for four different cases and show the results in Figure 1. As a reference, we use a  case with no extra radio background ($A_{\rm r} =0$) and refer to it as the ``standard'' astrophysical scenario. The other three examples  include a radio background  with amplitudes $A_{\rm r} = 1,~5.7,~18.4$, and different  spectral indexes:  $\beta = -2$, $-2.6$ and $-3$.  Other than $A_{\rm r}$ and $\beta$ the astrophysical parameters are fixed: $f_* = 3\%$, $f_X = 1$, $V_c = 16.5$ km/s, R$_{\rm mfp} = 30$ Mpc, $\zeta = 14$, hard X-ray SED with $\alpha = -1.3$, $\nu_{\rm min} =1$ keV. Even though the assumed radio background is uniform, it affects the fluctuations in the 21-cm signal by affecting $x_\alpha$,  $x_C$, $x_{\rm e}$ and $T_{\rm K}$. 

In our simulations, we assume that this background exists throughout cosmic history, and so it has a large effect on the 21-cm signal (both global signal and the power spectrum) from the dark ages, cosmic dawn and reionization. Its existence during the dark ages would signal the presence of exotic sources (e.g., dark matter annihilation or neutrino decay). If the extra background is generated by stars or stellar remnants, it will only affect the signal from cosmic dawn and reionization leaving the dark ages signature unchanged. Even though the selected form of the background may not be fully realistic, Figure 1 demonstrates that changing the spectral index and/or $A_{\rm r}$ leads to quantitative but not a strong qualitative difference in the signal. Although the dark ages and reionization signals rely on the assumption of constant $A_{\rm r}$, we expect to see a similar big picture in cases with a redshift-dependent amplitude and a more complicated spectrum. Moreover, as we discuss below,  $A_{\rm r}$ consistent with the EDGES Low-Band detection will saturate the dark ages signal for $\beta = -2.6$ or steeper spectral indexes leading to  model-independent predictions for both the global signal and the power spectrum.

\subsubsection{Dark Ages}

For the selected astrophysical model  the first population of stars forms around $z\sim 30$. At higher redshifts no stellar Ly-$\alpha$ or X-ray photons are present and the only couplings that affect the 21-cm line are due to the interaction with the background radiation or via intra-atomic collisions. In the case with $A_{\rm r} = 0$ the expected global signal has a weak absorption feature  at $z\sim 87$ of $T_{21} \sim -37$ mK and nearly vanishes at lower and higher redshifts due to the CMB coupling.  The power spectrum of fluctuations during the dark ages is imprinted by inhomogeneous density and velocity fields as well as fluctuations in $x_C$, $T_{\rm K}$ and $x_{\rm e}$ and is shown in the right panel of Figure 1   as a function of redshift at $k = 0.1$ Mpc$^{-1}$. In the standard case the power spectrum peaks around $z\sim 50$. The only uncertainty in the value of $T_{21}$ from the dark ages arises from the uncertainty in cosmological parameters (which we do not include in our estimates here). An extra radio background changes the global signature and the power spectrum dramatically. 

For the values of $A_{\rm r}$ and $\beta$ that we  show here the effect of the radio background is saturated (for   $A_{\rm r} = 5.7$ and $ \beta=-2.6$ and for  $A_{\rm r} = 18.4$ and $ \beta=-3$), or nearly saturated (for $\beta = -2$ and $A_{\rm r} =1$), during the dark ages. This can be understood by expressing $T_{21}$ to linear order in $T_{\rm rad}^{-1}$. For the synchrotron-like background that can explain the EDGES Low-Band detection,  $T_{\rm rad}$ is very large compared to other relevant temperatures  at low radio frequencies corresponding to the dark ages,  and as a result  $x_C<<x_{\rm rad}\sim 1$ (and in the absence of star formation $x_\alpha =0$).   We get $T_{\rm rad}T_{\rm S}^{-1}\sim 1+ x_CT_{\rm rad}/T_{\rm K}$. Ignoring the small effect of the radiative heating and because  $x_C\propto T_{\rm rad}^{-1}$, $T_{\rm rad}$ cancels from the expression  and we  obtain a  clear prediction for the 21-cm signal from the dark ages which does not depend on either $A_{\rm r}$ or $\beta$ 
\begin{equation}
T_{21}^{\rm lim}  = -\mathcal{N}\frac{n_i\kappa^i_{10}n_{\rm H}}{(1+z)^2 dv/dr}\frac{T_*}{T_{\rm K}},
\label{Eq:Tlim}
\end{equation}
with $\mathcal{N} = 3h_{\rm pl}c \lambda_{21}^2\left[32\pi k_B\right]^{-1}$.  If indeed such an extra radio background exists during the dark ages, the global signal would be  enhanced  by a factor of  $\sim 6$  at   $z \sim 87$,  and the power spectrum  by a factor of $\mathcal{O}(200-300)$ at the same redshift. 
 Owing to the uniquely predicted redshift dependence of a radio-saturated 21-cm signal (Eq. \ref{Eq:Tlim}), the nature of the EDGES signal can be easily identified if such a signal is observed, and the strong signal could be used for cosmology.

\subsubsection{Cosmic Dawn and Reionization}
   
For the adopted radio spectrum the intensity is lower at higher frequencies (lower redshifts) and the background is not strong enough to saturate the signal in most of the cases. As a result, astrophysical parameters have an effect and we recover the expected structure of peaks and troughs but with an enhanced or suppressed intensity (Figure 1 and 2);  moreover, features are shifted to lower redshifts. The effect of  $A_{\rm r}$ is highlighted  in Figure 2 for two cases: molecular and atomic cooling with fixed astrophysical parameters (see the caption of Figure 2 for details). Importantly, because the radio background does not produce any feedback on star formation,  astrophysical processes  (e.g., the build up of the Ly-$\alpha$  background) are independent of $A_{\rm r}$. However, the value of the coupling coefficient $x_{\rm tot}$ (Eqs. \ref{eq:xa}, \ref{eq:xc}) is lowered by $T_{\rm rad}$, and, thus, the coupling (defined as $x_{\rm tot} =1$) is delayed.  In addition, the radiative heating is changed, i.e., the excess radio background results in a higher temperature of the gas (and slightly higher $x_e$). However, because this effect is small compared to the heating by astrophysical sources, the heating transition (observationally defined as the redshift at which gas heats up to the brightness temperature of the radio background) is still delayed. 

\begin{figure*}
\begin{center}
\includegraphics[width=3.5in]{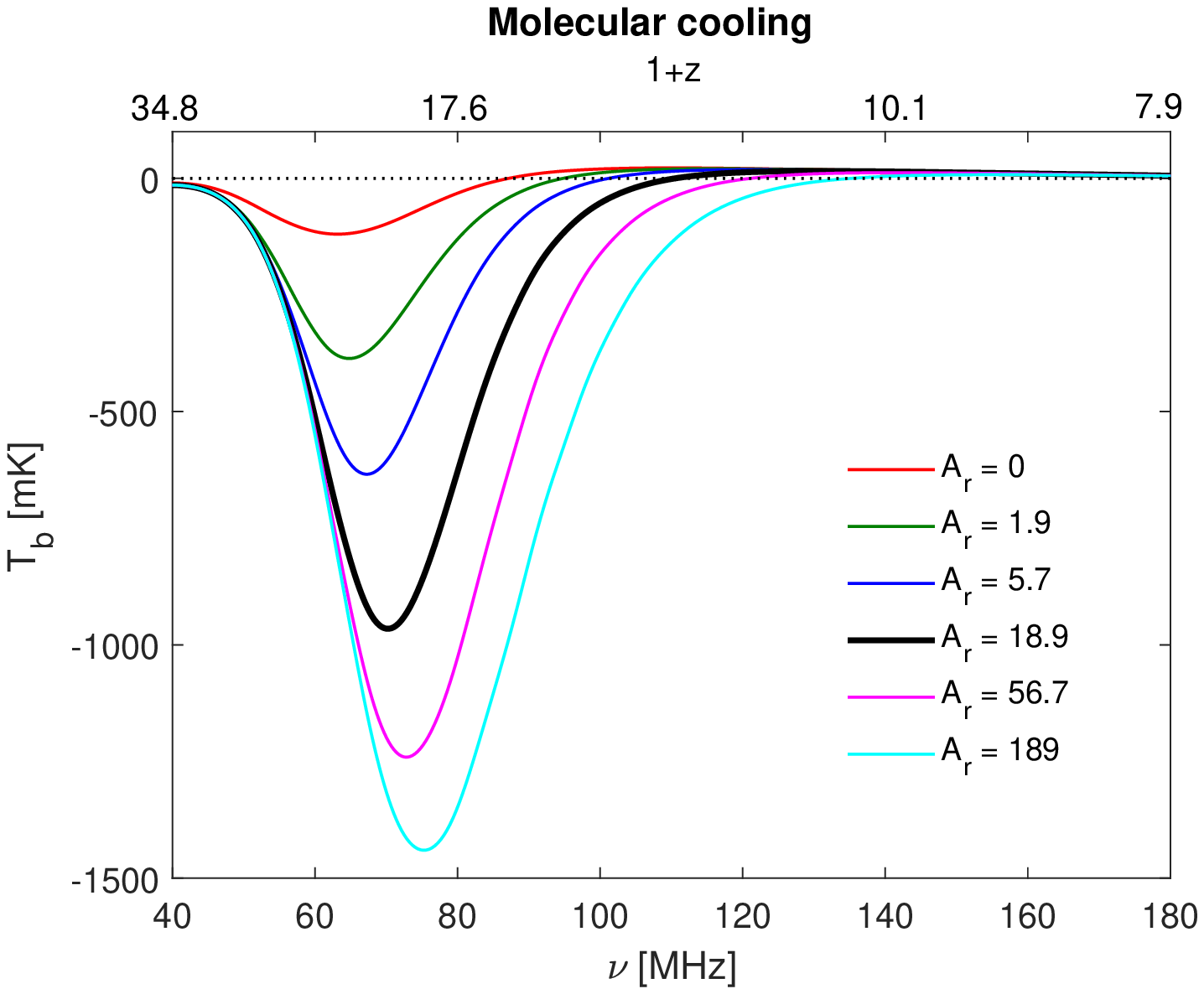}\includegraphics[width=3.5in]{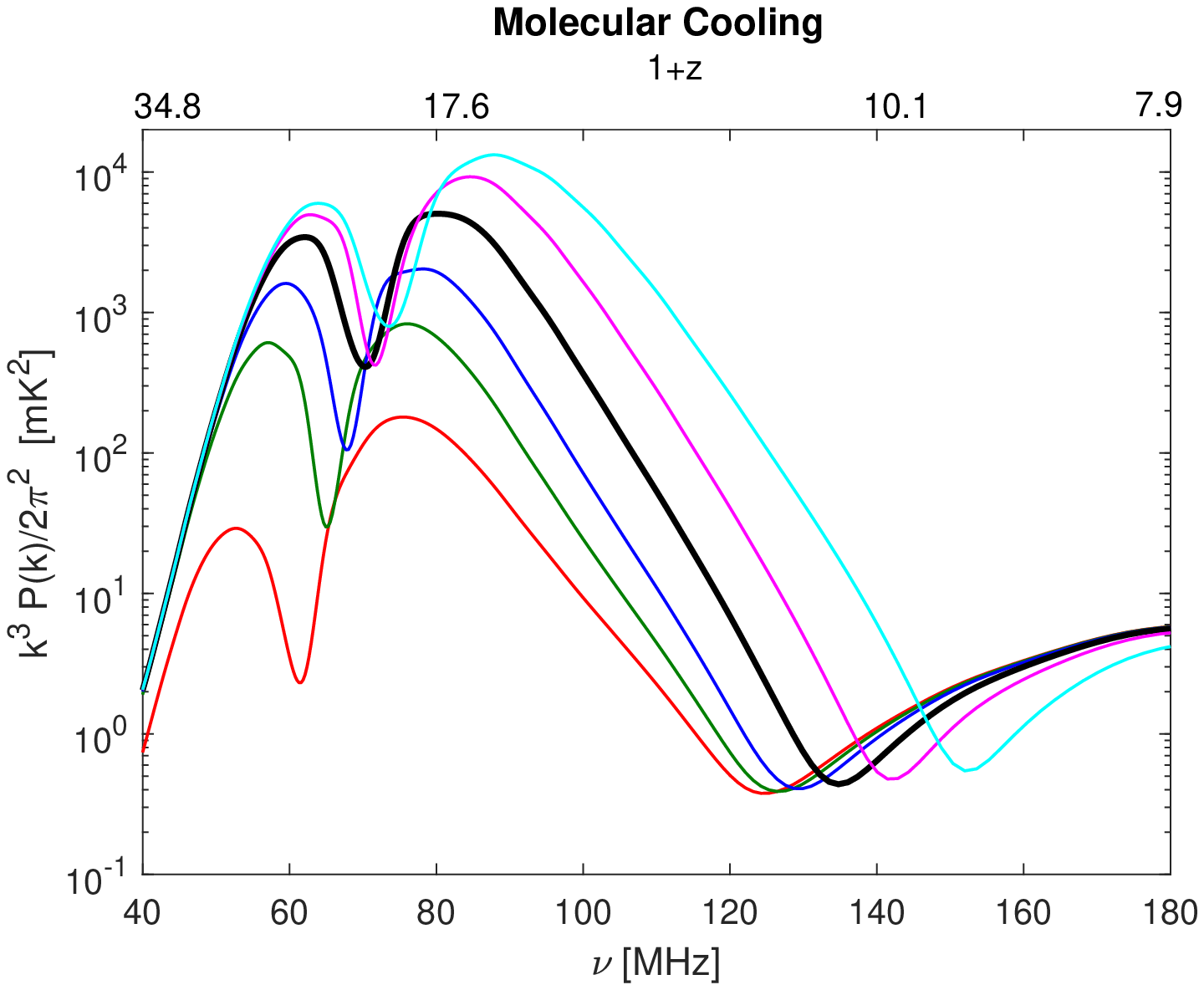}
\includegraphics[width=3.5in]{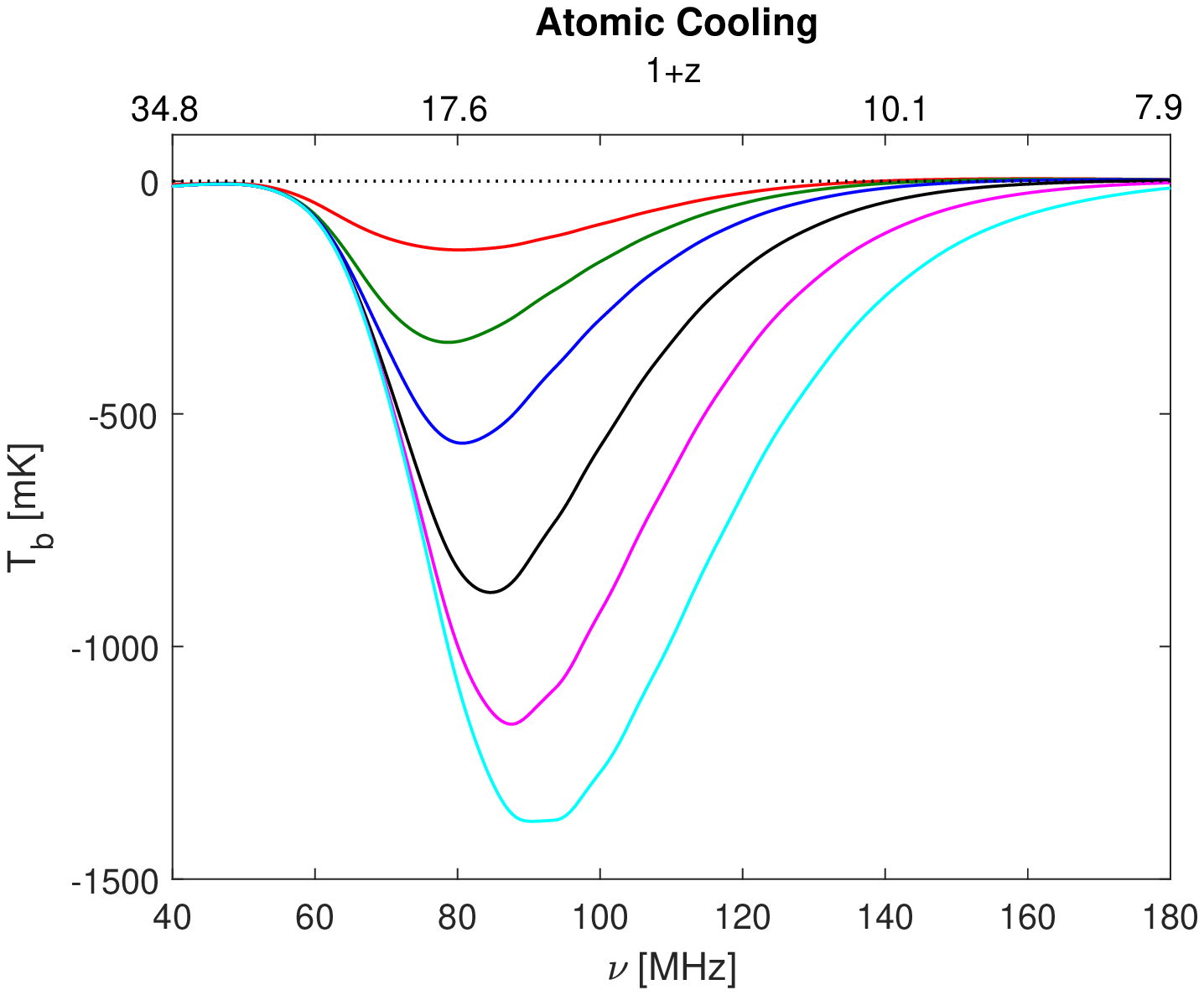}\includegraphics[width=3.5in]{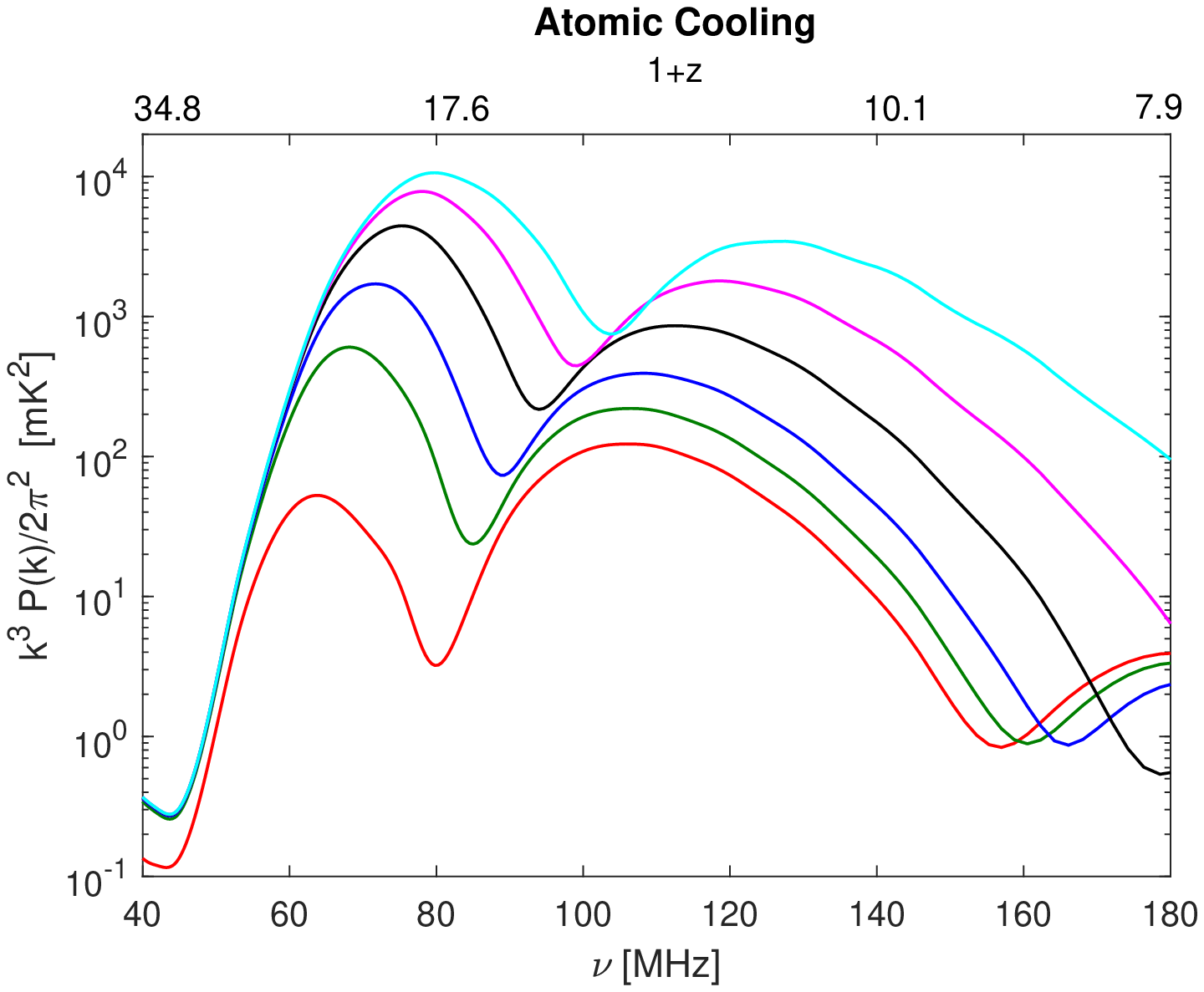}
\caption{Dependence of the cosmic dawn and reionization signals on  $A_{\rm r}$. We show the global signal (left) and the power spectrum at $k = 0.1$ Mpc$^{-1}$ (right) for the case of molecular cooling halos (top) with parameters $f_* = 10\%$, 
$f_X = 1$, $V_c = 4.2$ km s$^{-1}$, $\alpha = - 1.5$, $\nu_{\rm min} =0.1$ keV, $R_{\rm mfp}=30$ Mpc, $\zeta = 11.4$ (which yields $\tau  = 0.062$); and atomic cooling halos (bottom) with  parameters $f_* = 3\%$, 
$f_X = 0.1$, $V_c = 16.5$ km s$^{-1}$, $\alpha = - 1.5$, $\nu_{\rm min} =0.1$ keV, $R_{\rm mfp}=30$ Mpc, $\zeta = 14.2$ (which yields $\tau  = 0.064$)
 The cases are shown for $A_{\rm r} = 0$ (red), $A_{\rm r}= 1.9$ (green),   $A_{\rm r} = 5.7$ (blue),  $A_{\rm r} = 18.9$ (black),   $A_{\rm r} = 56.7$ (magenta),  $A_{\rm r} = 189$ (cyan). The case of molecular cooling halos with  $A_{\rm r} = 18.9$ (thick black curve, top) is consistent with all the observational constraints discussed in the next Section (\ref{Sec:params}).   } 
\end{center}
\label{fig:withAr}
\end{figure*} 

By design, at cosmic dawn the excess radio background generates strong absorption (left panels of Figure 2). For larger  $A_{\rm r}$ the absorption feature is deeper, making it easier to  observe. However, it is also  wider and slightly shifted to higher frequencies (lower redshifts). A corresponding change in the power spectrum is observed  (right panels of Figure 2). The latter is driven by  the effect of $T_{\rm rad}$ on the overall global intensity; however, the radio background also separately affects fluctuations in $x_\alpha$,  $x_C$, $T_{\rm K}$ and $x_e$. As a result of stronger  $A_{\rm r}$ the Ly-$\alpha$ and X-ray peaks are higher and shifted to lower redshifts (higher frequencies). 

The same radio background leads to a modification  of the reionization signature both in the global signal and in the power spectrum.
Cases which with weak $A_{\rm r}$  are seen in emission might appear in absorption for the stronger excess radio background (e.g., red versus cyan curve in Figure 2)  and the overall intensity of the 21-cm signal might be reduced.  This is because the redshift at which $T_{\rm S}$ reaches $T_{\rm rad}$ (and the signal transits from absorption to emission) is shifted to lower redshifts (deeper into the epoch of reionization).  In other cases, e.g., of inefficient heating, an excess radio background tends to enhance the absorption signal during reionization.

\section{Constraining astrophysical parameters with EDGES Low-Band}
\label{Sec:params}

We use the code outlined in Section \ref{Sec:Methods} to survey the space of astrophysical parameters, varying the parameters relevant for cosmic dawn and reionization, namely $V_c$, $f_*$, $f_X$, the X-ray SED, R$_{\rm mfp}$ and $\tau$ as well as the value of $A_{\rm r}$, but fixing the spectral index to that of the synchrotron emission $\beta = -2.6$.  We randomly probe the parameters, generating a set of 18309 different models, out of which 17002 models comply with the LWA1 limit. 

To verify the consistency of the models with EDGES Low-Band data we follow the approach we took in \citet{Fialkov:2018} by requiring the signal  to be deep, and localized within the band. Within $99\%$ confidence, the cosmological signal should
satisfy 
\begin{eqnarray}
 300~ \rm mK < && \bigl\{ \rm max \left[T_{21}(60<\nu<68)\right] \nonumber \\
&&  -\rm min\left[T_{21}(68<\nu<88)\right]\bigr\}  < 1000~\textrm{mK},
\label{Eq:C2}
\end{eqnarray}
and 
\begin{eqnarray}
 300~ \rm mK < && \bigl\{ \rm max\left[T_{21}(88<\nu<96)\right] \nonumber \\
&& -\rm min\left[T_{21}(68<\nu<88)\right]\bigr\}  < 1000~\textrm{mK}.
\label{Eq:C3}
\end{eqnarray}
In addition, we require consistency with the EDGES High-Band data \citep{Monsalve:2017} which rule out models with large variations in the  $100-150$ MHz band and imply
\begin{equation}
|T_{21}(100\, \rm{MHz})-T_{21}(150\, \rm{MHz})|<300~\textrm{mK}.
\label{Eq:C1}
\end{equation}
In this way, we conservatively demand only rough,
overall agreement with the current data. We also assume that $x_{\rm HI}(z=5.9)<20.7$\%, which is within $3\sigma$ of the value extracted using the dark fraction of QSO spectra \citep{Greig:2017}; and $x_{\rm HI}(z=7.5)<71.5$\% which is within $3\sigma$ of the neutral fraction measurement extracted from the damping wings \citep{Greig:2018} of ULASJ1342+0928, a bright QSO at z = 7.54 \citep{Banados:2018}. Finally, we require the extra radio background to be lower than or equal to the observed excess. An example of the signal consistent with all the observational constraints is shown in Figure 2 (thick black curve, top panels).   Out of 17002 examined models with $A_{\rm r}$ lower than the LWA1 limit, 1354 models pass the requirement of EDGES Low-Band data alone, while  686 models pass all the above constraints. The two  sets of the global signals and corresponding power spectra  are shown in Figure 3.

\begin{figure*}
\begin{center}
\includegraphics[width=3.5in]{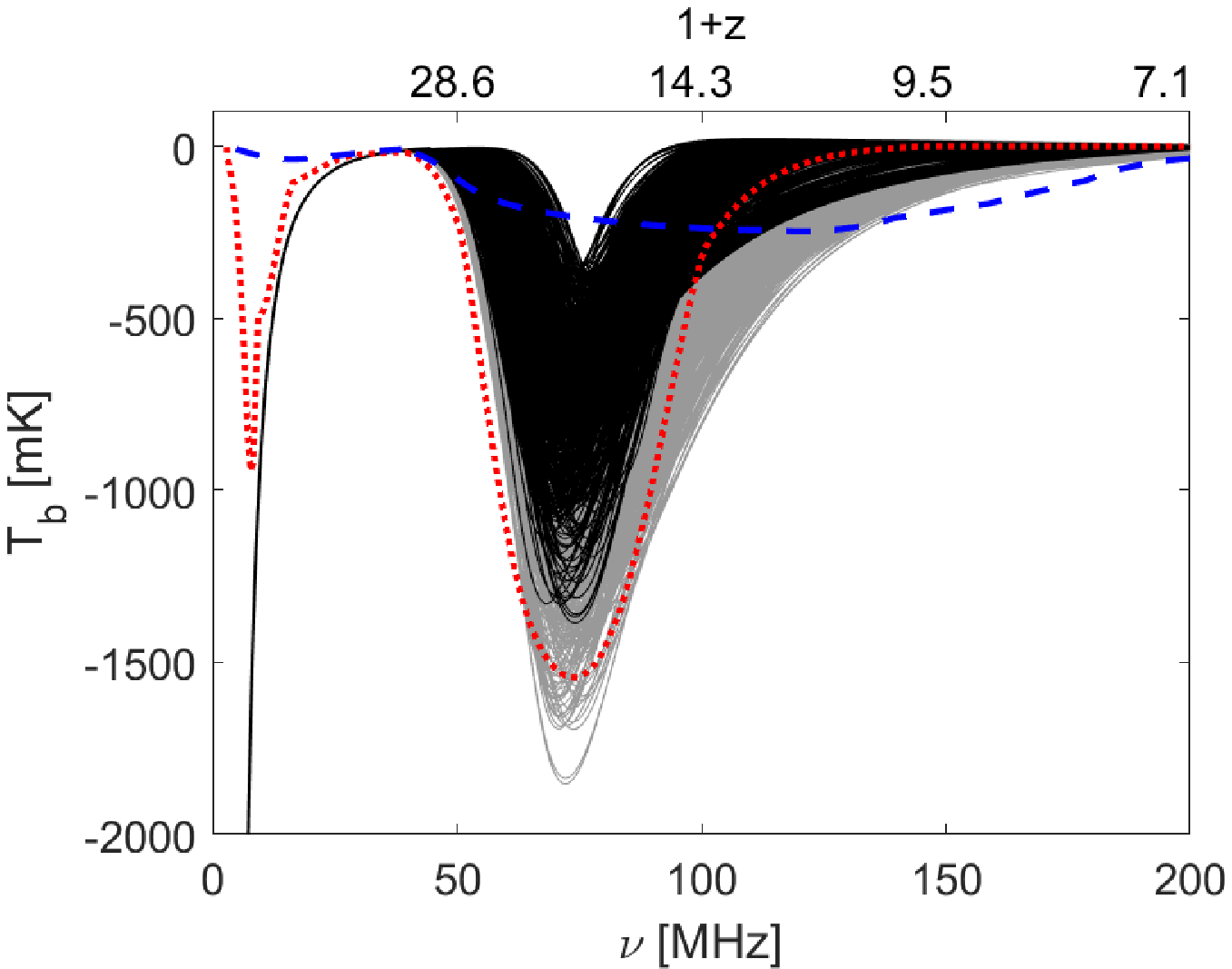}\includegraphics[width=3.5in]{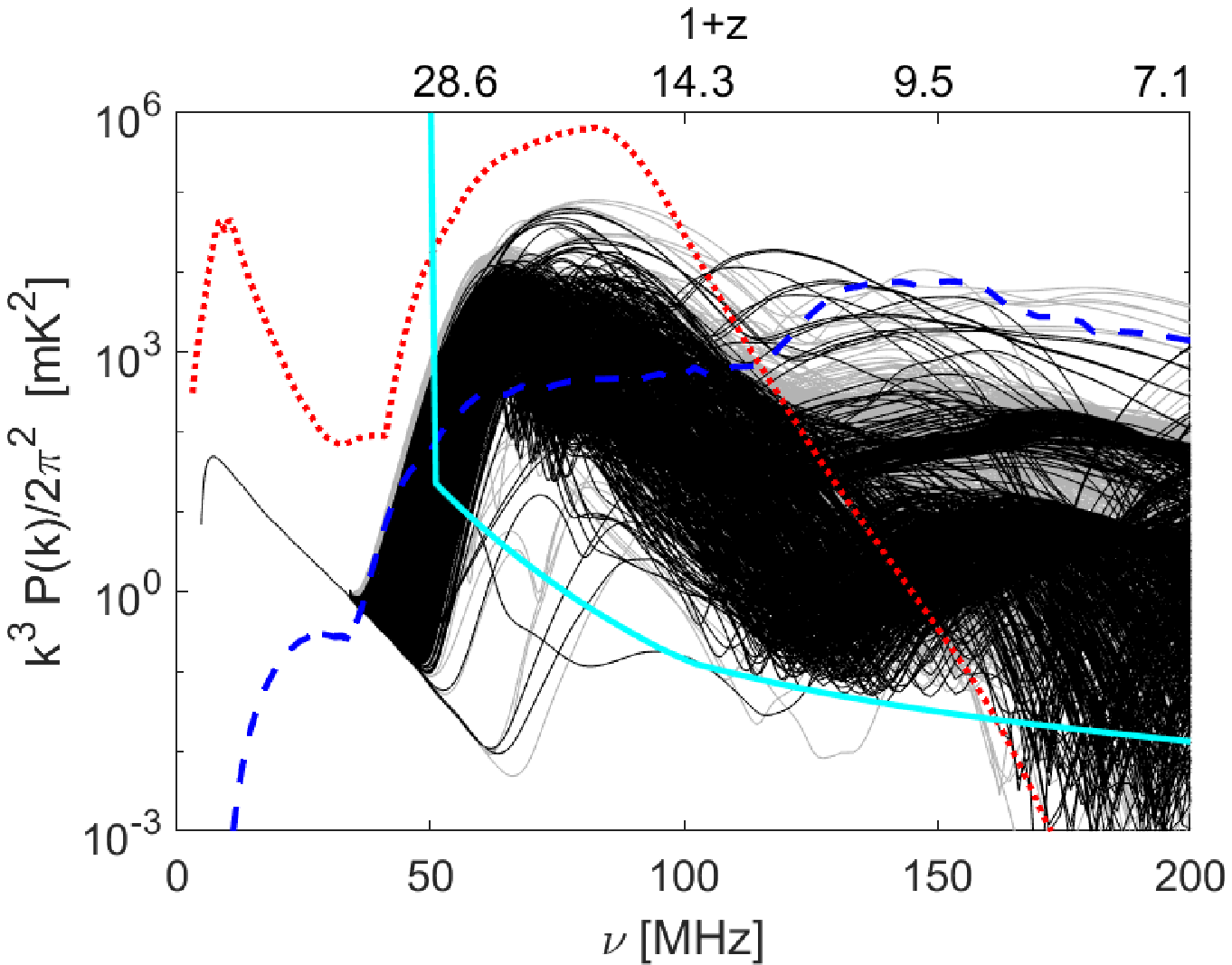}
\caption{Summary of models. We show the global signal (left) and the power spectrum at k = 0.1 Mpc$^{-1}$ (right)  vs frequency/redshift (bottom/top). Models that are compatible with EDGES Low-Band and are also  allowed by the LWA1 data are shown in grey (1354 cases); models compatible with EDGES Low-Band, High-Band, high-redshift quasars at $z = 5.9$ and $z=7.5$ and are also  allowed by the LWA1 data  are shown in black (686 cases); the upper limit on $T_{21}$ and on the power spectra for models which do not include the extra radio background ($A_{\rm r} = 0$) is shown in dashed blue.  The dotted red line shows the upper limit for models with b-dm  scattering which are discussed in Section \ref{Sec:disc} \citep[red, valid models in][]{Fialkov:2018}; we note
that this line includes fluctuations only due to the b-dm scattering
and does not include astrophysical sources of fluctuations. On the power spectrum plot we also show the SKA1 noise curve (solid cyan). } 
\end{center}
\label{fig:signalsAll}
\end{figure*} 

As we demonstrated in Section \ref{Sec:case}, for $A_{\rm r} = 5.7$ and  $\beta = -2.6$ the effect of the radio background is saturated during the  dark ages, and there is no uncertainty in both the predicted power spectrum and the global signal. As we can see in Figure 3, this remains valid for all the values of $A_{\rm r}$ allowed by the data, $1.9<A_{\rm r}<418$. Therefore, if the anomalous EDGES Low-Band signal is created by exotic physics via an excess radio background, the dark ages could potentially confirm this origin (for models in which the radio
background was present at such early times). However, at lower redshifts, during cosmic dawn and reionization,  the allowed models show a lot of variety as a result of the interplay between  the radio background and other astrophysical parameters. Both the global signals and the power spectra are boosted at $z\sim 17$, with an enhancement in the squared fluctuation of up to $\mathcal{O}(100)$ compared to the strongest allowed signals with $A_{\rm r} =0$ (the dashed blue curve in Figure 3 shows  the envelope of all the cases with no radio background for comparison). During reionization the signals that with $T_{\rm CMB}$ would be seen in emission  are mostly suppressed because the stronger background radiation leads to a weaker contrast between $T_{\rm rad}$ and $T_K$. Even with the suppression, however, all the considered scenarios are within reach of the Square Kilometer Array \citep[SKA, the predicted power spectrum sensitivity curve is shown in cyan,][]{Koopmans:2015}.

Existing observations allow us to constrain astrophysical parameters of cosmic dawn. In Figure 4 we show the values of astrophysical parameter for the  686 models consistent with all the observational constraints (black circles; complying with EDGES Low-Band, EDGES High-Band, quasars, and LWA1 constraints) and the  1354 models consistent  with the EDGES Low-Band and LWA1 data (grey crosses)  along six different cuts of the explored parameter space, namely: $f_*-f_X$, $f_*-V_c$, $f_X-V_c$, $A_{\rm r}-f_X$, $A_{\rm r}-V_c$, $A_{\rm r}-f_*$. Below we summarize the astrophysical constraints.

\begin{figure*}
\begin{center}
\includegraphics[width=2.2in]{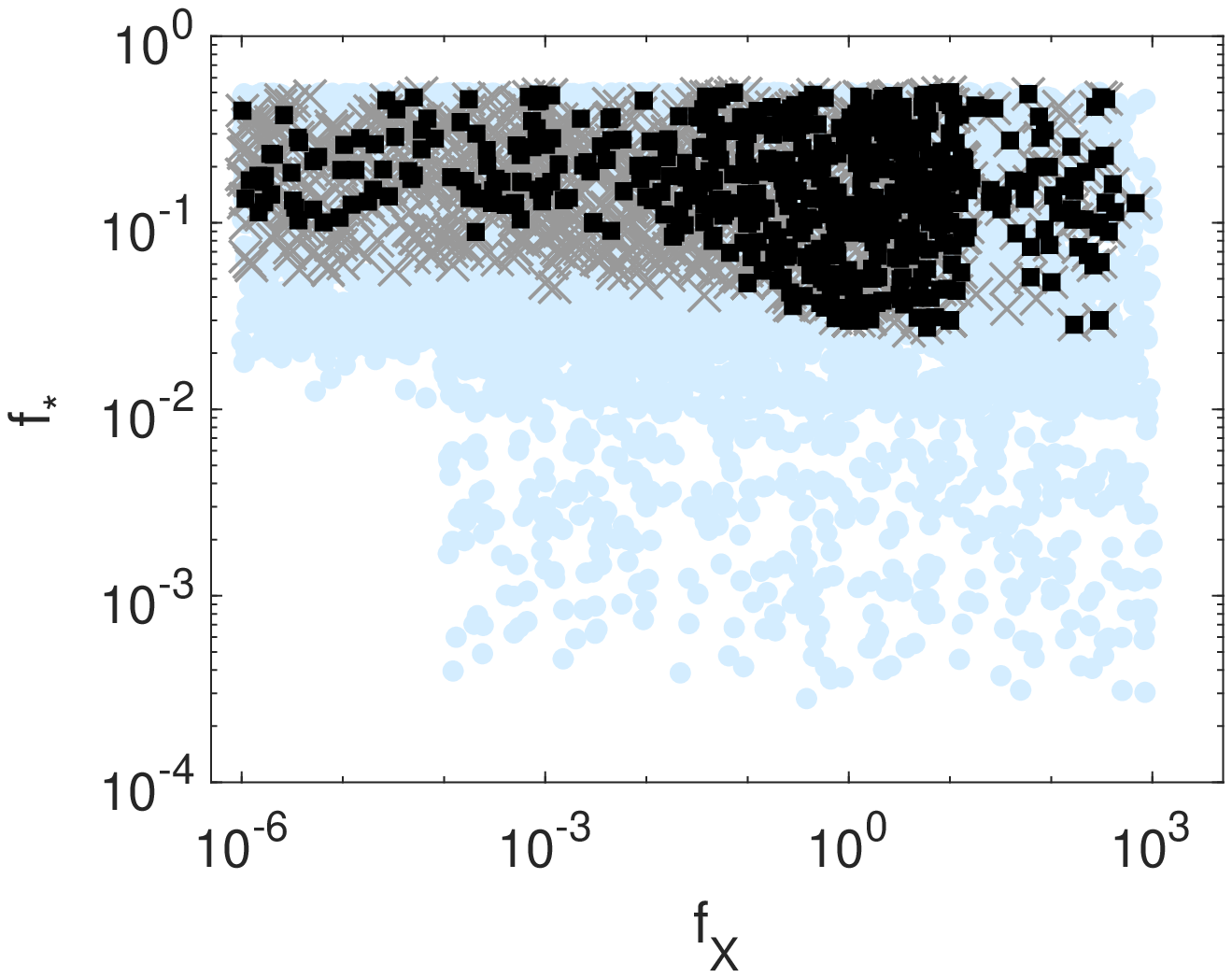}\includegraphics[width=2.2in]{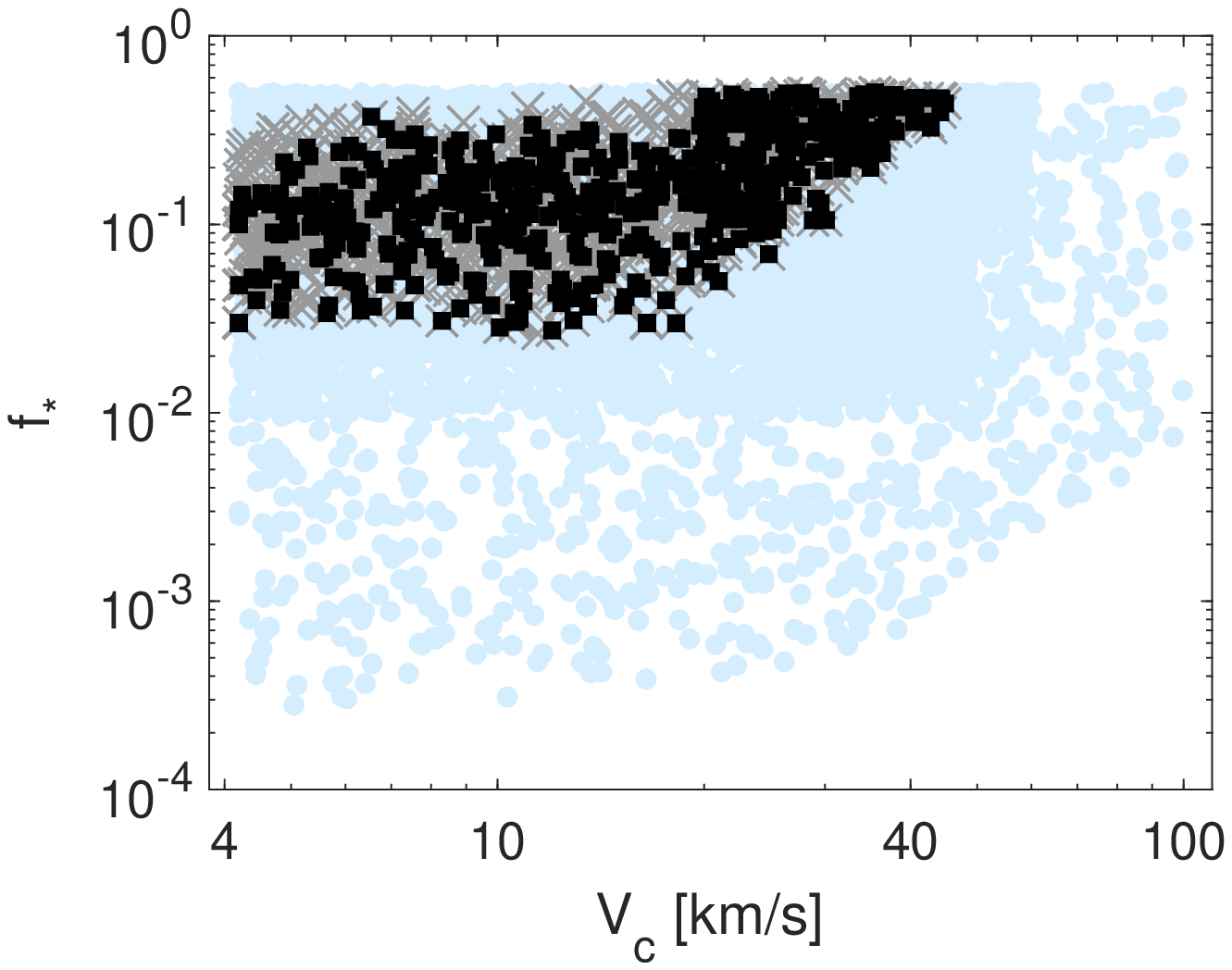}\includegraphics[width=2.2in]{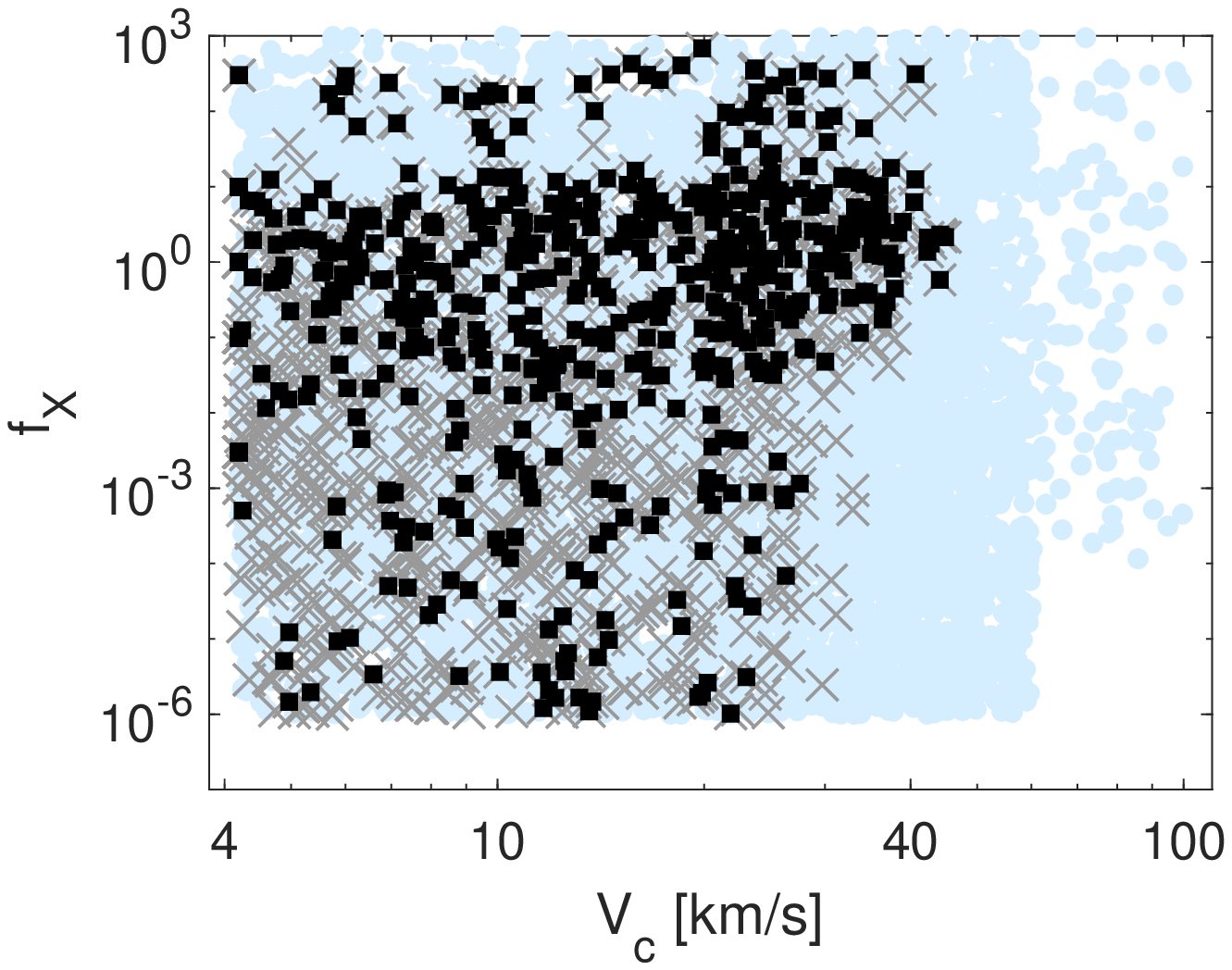}
\includegraphics[width=2.2in]{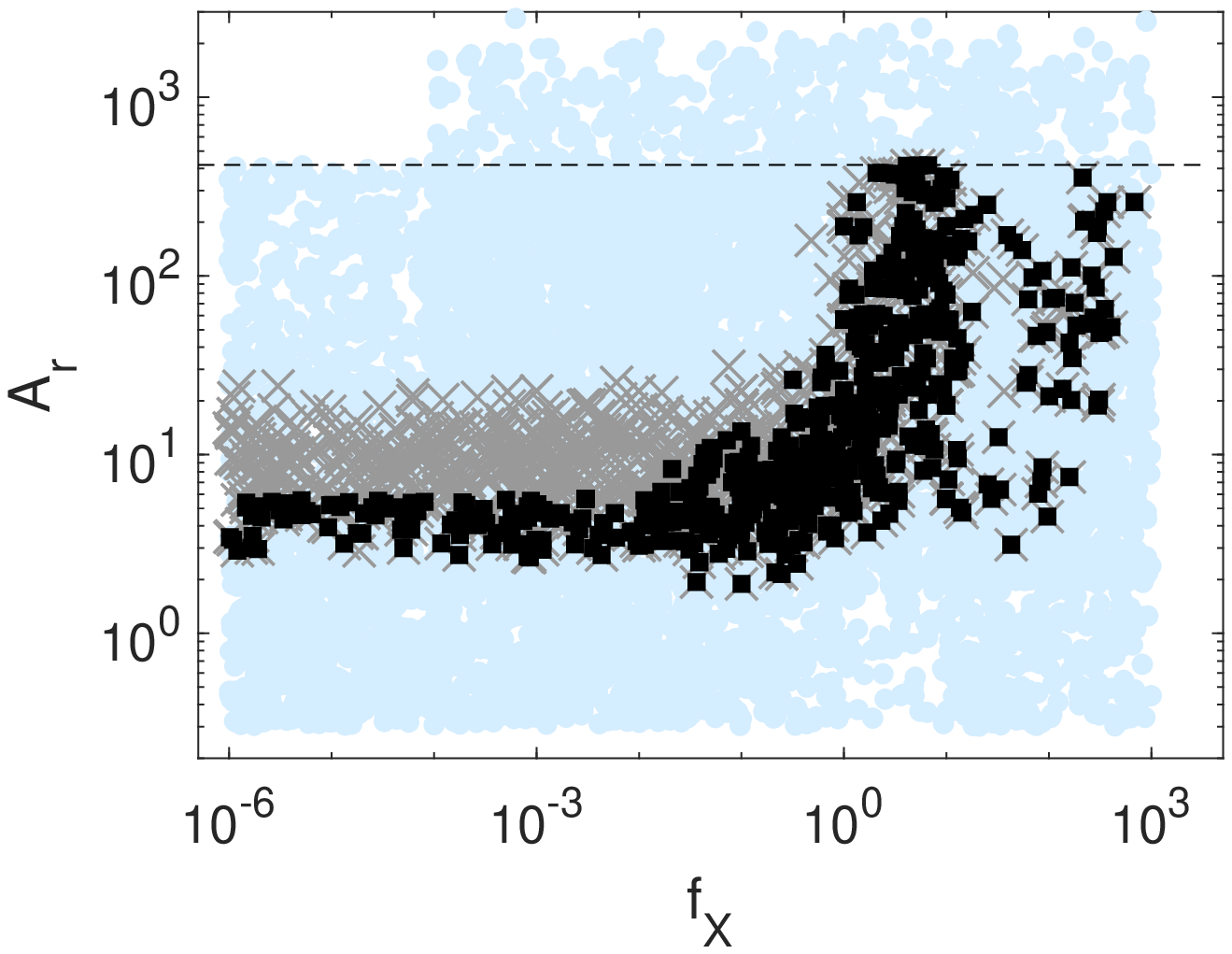}\includegraphics[width=2.2in]{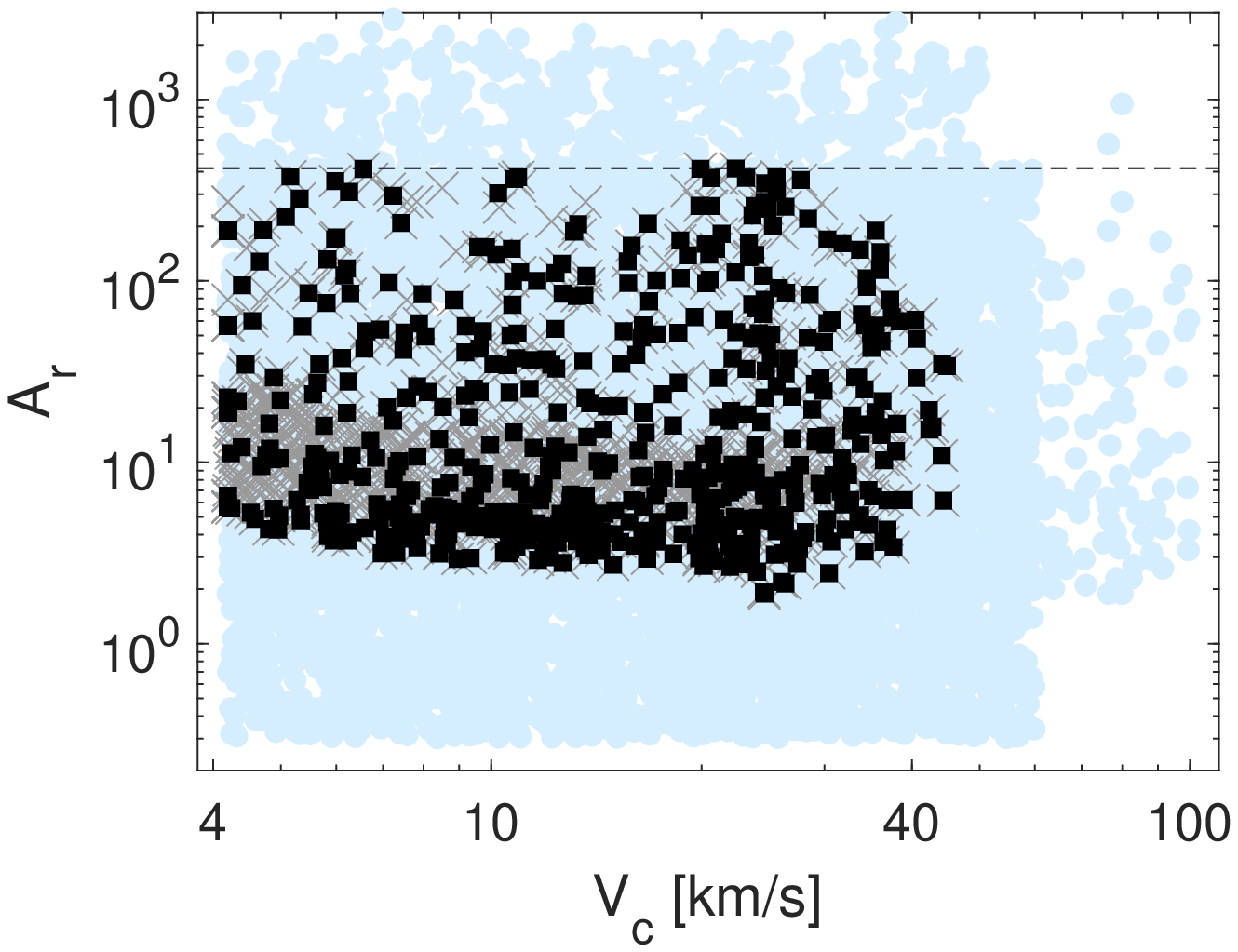}\includegraphics[width=2.2in]{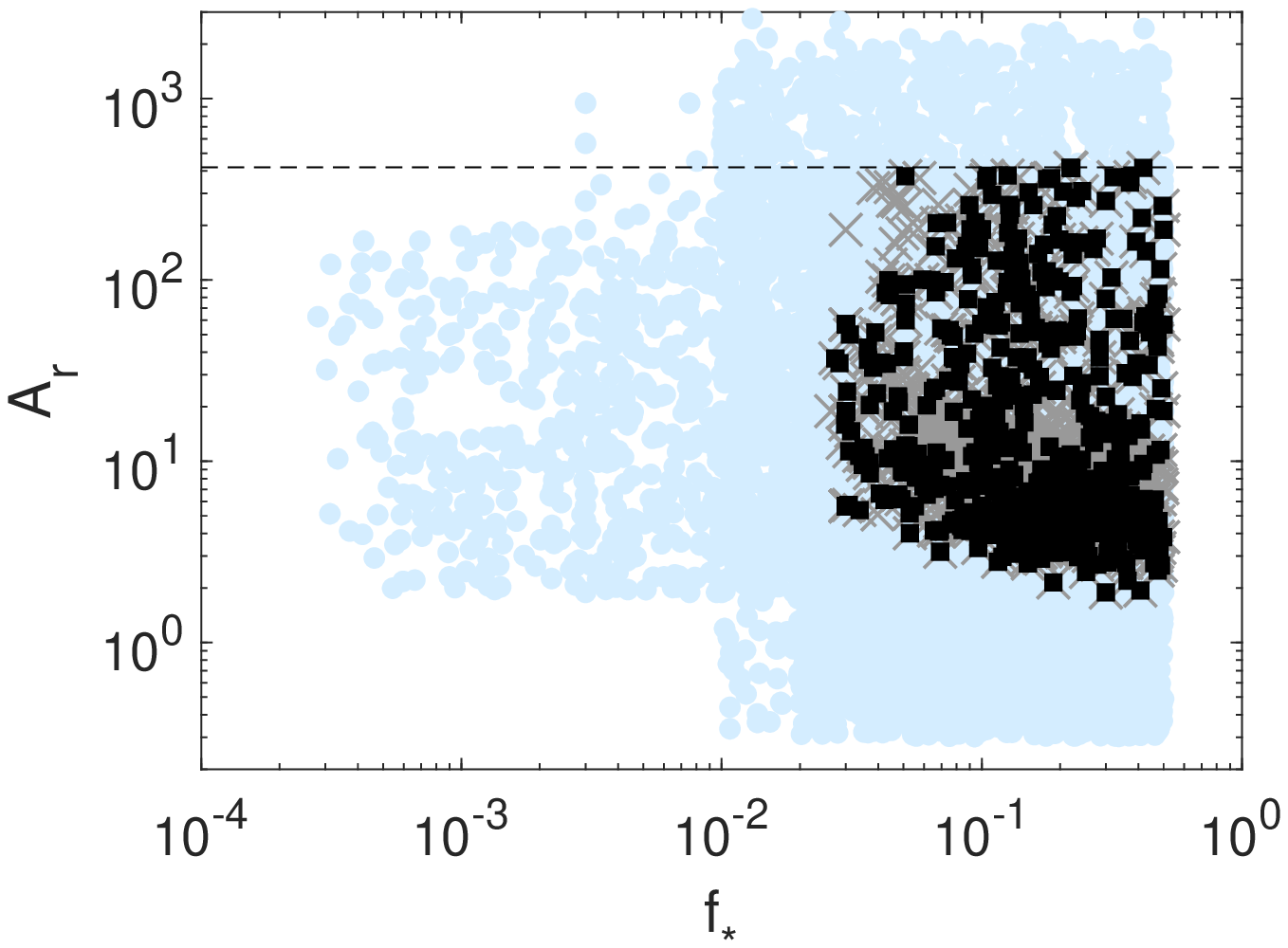}
\caption{ Astrophysical parameter space of allowed models that are compatible with EDGES Low-Band data and the LWA1 data (1354 cases, grey crosses); models compatible with EDGES Low-Band, High-Band, high-redshift quasars at $z = 5.9$ and $z=7.5$ and the LWA1 data (686 cases, black dots). Blue dots demonstrate the  parameter space surveyed in this study as specified in Section \ref{Sec:astro}, in total 18309 models  were generated out of which 90\% of randomly selected models are shown. The dashed line (bottom panels) shows the LWA1 limit.}
\end{center}
\label{fig:params}
\end{figure*}

\subsection{Excess Radio Background}
Our analysis shows that the models consistent with the above-mentioned constraints share high values of the amplitude of the radio background, $A_{\rm r}>1.9$ at the reference frequency of 78 MHz (equivalent to  0.1\% of the CMB at the  reference frequency of 1.42 GHz). Out of the explored range  $0.3<A_{\rm r}<2000$, the values between  1.9  set by the EDGES Low-Band detection and the absolute upper limit $A_{\rm r,max} = 418$ set by the LWA1 observation are allowed.  However, there is a degeneracy between the values of $A_{\rm r}$ and $f_X$  (lower left panel in Figure 4) because the two parameters affect the depth of the absorption feature. We find that for values of $f_X$ much higher than unity  for which the absorption trough in the absence of the radio background is very shallow \citep[the shallowest absorption trough found in the recent parameter study by][is of $-25$ mK]{Cohen:2017}, higher values of  $A_{\rm r}$ are allowed by the EDGES-Low data  with the maximal value fixed by the LWA1 observation and a slightly elevated lower limit of $\sim 4$. For low values of $f_X$ the absorption trough is deep even for $A_{\rm r} = 0$  \citep[the deepest absorption trough found in the parameter study by][is of $-240$ mK]{Cohen:2017}. Therefore,  the maximum allowed  $A_{\rm r}$ decreases as $f_X$ decreases. For $f_X \ll 1$ X-ray sources have a very weak effect on the 21-cm signal, and the allowed range of $A_{\rm r}$ is independent of $f_X$: $1.9< A_{\rm r} \sim 25$ with EDGES Low-Band constraint only and $1.9< A_{\rm r} \sim 5.8$ when the lower redshift constraints are included.  (Note that the limits that we derive here depend on the spectral index. The values of $A_{\rm r}$ are expected to be higher for more negative $\beta$ in agreement with the trend in Figure 1.)

The minimum required value of $A_{\rm r} = 1.9$ may seem somewhat high
compared to the value $\sim 1$ expected based simply on the minimum
absorption depth measured by EDGES. The reason is that the EDGES
measurement gives a combination of constraints, requiring the high-redshift feature to be deep and narrow at the same time. 
The narrowness of the
measured absorption feature, and the limited frequency range of the
entire EDGES measurement, imply that in some models only part of the
total absorption feature is captured within the EDGES-Low band. In these cases the entire profile is wider and deeper than the reported signal,  and higher values of $A_{\rm r}$ are needed to explain the observed contrast.

\subsection{X-ray Heating Efficiency}

It was recently shown that,  in the absence of an excess radio background,  non-detection of the 21-cm signal by EDGES High-Band places a lower limit of  $f_X = 0.0042$      \citep{Monsalve:2018b}. Based on the compilation of models examined in the present paper we find that, owing to the contribution of the excess radio background,  no values of $f_X$ are ruled out by the low-redshift data\footnote{However,  only a qualitative comparison can be made here with the results of \citet{Monsalve:2018b} where a much more rigorous data analysis and joint fitting of the modeled 21-cm signals and foregrounds to the data were performed. }
 
We find that the data  do not constrain $f_X$, and  the only condition on $f_X$ is imposed by the degeneracy with $A_{\rm r}$ described above: at values of   $A_{\rm r}$ higher than $\sim 25$ only $f_X\gtrsim 1$ are allowed because the lower values of $f_X$ would generate steeper (and deeper) than observed signals. At lower values of $A_{\rm r}$ lower values of $f_X$ are allowed (in agreement with the lower left panel of Figure 4). Finally, at the lowest possible  $A_{\rm r}$, $f_X\gtrsim 0.1$ are excluded as they would create a shallower than detected absorption feature. 

Our conclusion is different from the results of \citet{Mirocha:2018} who placed a lower limit of $f_X=10$. This is because \citet{Mirocha:2018} required the signal to be entirely within the band and the total absorption to match the observed trough in absolute depth; while our Eqs. 11, 12, and 13 imply that in some cases only part of the total absorption profile is within the EDGES-Low band.
 In addition, \citet{Mirocha:2018} considered models in which the radio background is tied to the star formation history.

\subsection{Star Formation Efficiency and Mass of Star Forming Halos}

Ly-$\alpha$ coupling governs the high-redshift end of the cosmic dawn signal and, regardless of $f_X$, is constrained by the EDGES Low-Band observation which requires intense star formation before $z=17$. The efficiency and timing of the coupling are determined by $f_*$ and $V_c$ together.

 Using the non-detection of the 21-cm signal with the EDGES High-Band data, \citet{Monsalve:2018b} excluded intermediate values of $f_*$ between  $0.4\%$ and $3.9\%$ and found degeneracy in the $f_*-f_X$, $f_*-V_c$, and $V_c-f_X$ planes. The detection in the low frequency band adds a new constraint.  Out of the considered range,  $0.01\% < f_* < 50\%$, only $ f_*>2.8\%$ is consistent with the data. Lower values of $f_*$ result in a wide absorption feature which is shifted to lower redshifts and  would be outside the observed $50-100$ MHz band.   This lower limit is nearly independent of $f_X$ or $A_{\rm r}$, but strongly depends on $V_c$ at halo masses above the atomic cooling threshold. This is because for high values of $V_c$ the absorption trough shifts to lower redshifts and outside of the EDGES Low-Band and only very high values of $f_*$ can compensate for this trend and shift the signal back to higher redshifts.  Below atomic cooling, feedback limits the contribution of small halos. Also, considering the high- and the low-redshift constraints
together gives a dependence of the minimum value of $f_*$ with $f_X$.

We find that, out of the considered range  $4.2 < V_c < 100$ km s$^{-1}$ (or, equivalently $6.3\times 10^5<{\rm M}_{\rm h}<8.5\times 10^9$ M$_\odot$ at $z = 17$), the values of $V_c$ between  $4.2$ and $45$ km s$^{-1}$ (i.e., M$_{\rm h}<7.8\times 10^8$ M$_\odot$ at $z = 17$) are allowed. 
The maximum $V_c$ is set by our upper limit on $f_*$ of 50\%. If, e.g., we
were to only allow $f_*$ up to 10\%, the maximum $V_c$ would then be 30
km s$^{-1}$ ($2.3\times 10^8$ M$_\odot$ at $z=17$).  This limit is consistent with what is reported by \citet{Mirocha:2018} and \citet{Schauer:2019} who found that star formation in halos below $10^8$ M$_\odot$ is required to explain the EDGES Low-Band signal. In addition, non-detection of the signal by the EDGES High-Band  separately requires the existence of small star forming halos of  M$_{\rm h}<1.3\times 10^8$ M$_{\odot}$  \citep{Monsalve:2018b}.

 \section{Discussion}
 \label{Sec:disc}
Scenarios that we have explored here intend to explain the deep absorption signal detected by EDGES Low-Band via an excess radio background. Another set of models proposed in the literature as a solution to the EDGES Low anomaly are the models with non-gravitational baryon-dark matter scattering \citep{Barkana:2018}. Here we point out major differences between the two types of signature. 

In Figure 3 we show the envelope of all signals with b-dm scattering that comply with the EDGES data (red curves) which we explored in \citet{Fialkov:2018}.  This set of models explores the full astrophysical parameter space, to which dark matter mass and scattering cross-section were added. The models assume a velocity dependent cross-section of the form $\sigma\propto v^{-4}$ and that 100\% of dark matter  scatter off baryons. Even though scenarios in which 100\% of the dark matter interacts are ruled out by observations \citep{Barkana:2018b, Berlin:2018, Munoz:2018}, non-gravitational  scattering of baryons off dark matter  is still a plausible scenario assuming that only a few percent of dark matter bears electric charge of a few times $10^{-6}$ of the charge of an electron. Such scenarios can still be realized (with properly re-scaled cross-section) and it is still interesting to point out smoking gun signatures of the two classes of models. We also note that other particle
physics scenarios are still being explored \citep[e.g.,][]{Falkowski:2018, Houston:2018}.

In the absence of astrophysics, the dark ages provide a unique environment to constrain exotic physics. The global signal in the case of b-dm scattering  features a deeper trough  than the one expected from the standard case (i.e, $\sigma = 0$ and $A_{\rm r} = 0$, the blue line in Figure 3 shows the envelope of such models); while in the cases with $A_{\rm r} \neq 0$ the signal grows with increasing redshift (as long as the radio
background has already appeared). Because the radio background itself is assumed to be uniform, fluctuations are not as high as those that are possible in the b-dm scattering case where the power is boosted by the velocity dependent cross-section. 

At cosmic dawn   and reionization, the signatures of both the radio background and the b-dm scattering are contaminated by  astrophysical processes and it would be harder   to uniquely establish the nature of the enhancement in the global signal. However, the expected power spectra bear unique signatures and are key to identifying the nature of the enhancement. Because of the velocity-dependent cross-section, in the case of b-dm scattering the power spectrum is (1) enhanced by up to four orders of magnitude compared to the standard case, and  (2) bears a unique signature of BAO imprinted by the relative velocity between dark matter and gas, $v_{\rm bc}$ (see Section \ref{Sec:astro}). Examples of BAO at $z=17$ in the spectral shape, illustrated using the ratio of the power spectrum in mK$^{2}$ units ($k^3P(k)/2\pi$) to its value at $k=0.1$ Mpc$^{-1}$, are shown in Figure 5 as a function of the comoving wavenumber. For comparison, the enhancement in the power spectrum in the case of $A_{\rm r}\neq 0$ is weaker (only up to two orders of magnitude higher than the standard astrophysical scenarios, compared to four orders of magnitude in the b-dm case), and the spectral shape does not feature BAO. 
 
\begin{figure}
\begin{center}
\includegraphics[width=3.5in]{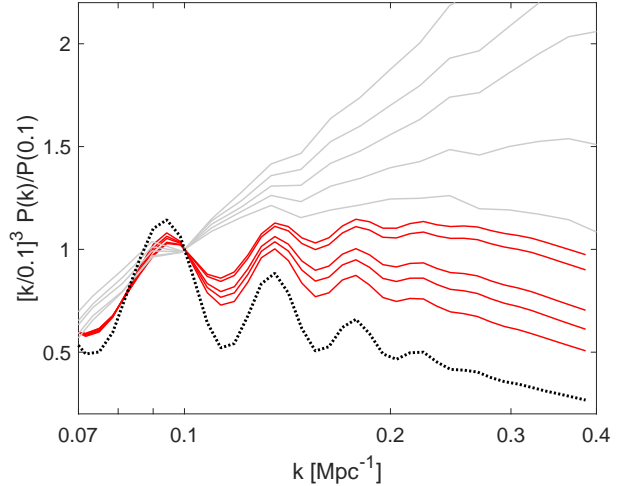}
\caption{Spectral shape with an excess radio background (grey) and b-dm scattering (red) at $z = 17$. A few
randomly-chosen models are shown to illustrate each category of
models. The shape of the power spectrum of the b-DM relative velocity is also shown (dotted black). } 
\end{center}
\label{fig:comp}
\end{figure}

\section{Summary}
\label{Sec:sum}

In this paper we have examined the effect of a uniform  radio background, existing in addition to the CMB, on the 21-cm signal from the dark ages, cosmic dawn and reionization, assuming a synchrotron-like spectrum of spectral index $\beta = -2.6$  and redshift-independent amplitude. We demonstrate both the global signal and the power spectrum computed in the $6-300$ redshift range. 
We vary conventional astrophysical parameters as well as the amplitude of the extra radio background and put limits on astrophysical parameters by requiring the broad consistency of models with the data of EDGES Low-Band, EDGES High-Band, limits on $x_{\rm HI}$ from quasars at $z=5.9$ and $7.5$ as well as the observed excess of the radio background detected by LWA1 between 40 and 80 MHz. Finally, we point out that  a non-uniform radio background would add a new source of fluctuations which are unaccounted for in present-day 21-cm codes.

The effect of an extra radio background is to deepen  and widen the absorption trough.  We find that the reported global signal at low frequencies is consistent with values of the amplitude  of the excess radio background of $1.9<A_{\rm r}<418$ relative to the CMB at the 78 MHz reference frequency of cosmic dawn (corresponding to $0.1-22\%$ of the CMB at 1.42 GHz). The   EDGES Low-Band detection requires  efficient star formation ($f_* > 2.8\%$) occurring in  small halos with  circular velocity below $V_c= 45$ km s$^{-1}$ (corresponding to M$_{\rm h}<7.8\times 10^8$ M$_\odot$ at $z = 17$). We find that for high $V_c$ higher values of $f_*$ are required.  This measurement alone does not place bounds on  $f_X$. The only constraint is that at high $A_{\rm r}$ higher values of  $f_X$ are required to fit the data. Adding the low-redshift data (quasars, EDGES High-Band) restricts the range of the allowed values of $A_{\rm r}$; however, contrary to what is found with the low-redshift data alone \citet{Monsalve:2018b}, there is no lower bound on $f_X$.   Finally, we find degeneracy in the $V_c-f_X$, $V_c-f_*$, and $V_c-f_*$ planes, in  broad agreement with the joint analysis by \citet{Monsalve:2018b} of the EDGES High-Band data and quasar and Lyman-break galaxies data. 

The 21-cm fluctuations during cosmic dawn are increased correspondingly to the global signal, while fluctuations from the epoch of reionization are either suppressed or enhanced depending on the parameters. Compared to the case of b-dm scattering which features strong BAO due to the velocity dependent cross-section, the spectrum in the case of the extra radio background has a smooth shape  and is not as strongly enhanced on large scales. Low frequency observations with interferometers such as  LOFAR, HERA and, once operational, the SKA, will be able to test the detection by the EDGES Low-Band and, if confirmed, assess  the nature of the enhancement via the magnitude of the fluctuations and the presence or absence of the BAO. 

Finally, we find that the signal from the dark ages  can be used to identify the origin of the EDGES Low-Band signal and
is the unique discriminator of the signature of the radio background (if the radio background is already 
present at such early times) versus scenarios such as baryon-dark matter scattering. If the extra radio background is realized in Nature  and is present prior to the onset of star formation, the dark ages signal will be enhanced.  We find    that any $A_{\rm r}$ that explains the EDGES Low-Band detection will saturate the dark ages signal for $\beta = -2.6$ or steeper spectral indexes. In this case the dark ages signal (both the global signal and the power spectrum) is   model-independent and has a unique redshift dependence. Several satellite  missions are being discuses to detect the 21-cm signal from the dark ages which will enable us to test the presence of the radio background at lower frequencies and, if the anomalous signature is detected, verify its origin \citep[e.g.,][]{Burns2019}.

\section{Acknowledgments}

We acknowledge the usage of the Harvard {\it Odyssey} cluster. Part of this project 
was supported by the Royal Society University Research Fellowship of AF. 
This project/publication was made possible for RB through the support
of a grant from the John Templeton Foundation. The opinions expressed
in this publication are those of the authors and do not necessarily
reflect the views of the John Templeton Foundation. RB was also
supported by the ISF-NSFC joint research program (grant No. 2580/17).

\end{document}